\documentclass[useAMS,usenatbib]{mn2e}

\usepackage{graphicx}
\usepackage{verbatim}
\long\def\symbolfootnote[#1]#2{\begingroup%
\def\thefootnote{\fnsymbol{footnote}}\footnote[#1]{#2}\endgroup}

\title[Stellar population gradients]{Stellar population gradients in early-type cluster galaxies\thanks{Based on observations from the European Organisation for Astronomical Research in the Southern Hemisphere, Chile (programmes: 078.B-0539, 081.B-0539)}}
\author[T. D. Rawle et al.]{T. D. Rawle\thanks{E-mail:
t.d.rawle@durham.ac.uk}, Russell J. Smith, J. R. Lucey\\
Department of Physics, Durham University, Durham DH1 3LE, United Kingdom}

\begin{document}

\date{Accepted 15 September 2009}

\pagerange{\pageref{firstpage}--\pageref{lastpage}} \pubyear{2009}

\maketitle

\label{firstpage}

\begin{abstract}
We present a study of internal stellar population gradients in early-type cluster galaxies. Using the VLT VIMOS integral field unit, we observed 19 galaxies in the core of the Shapley Supercluster ($z$ = 0.048). The radial trends in nine absorption lines (H$\delta$F to Fe5406) were measured to the effective radius for 14 galaxies, from which we derived the gradients in age, total metallicity and $\alpha$-element over-abundance. We combine these with results from 11 galaxies studied in our previous VIMOS work \citep{raw08-1891}. We observe a mean metallicity gradient of --0.13 $\pm$ 0.04 dex$^{-1}$ and, in common with the findings of previous studies, galaxies with log$\,\sigma$ $\ga$ 2.1 have a sizeable intrinsic scatter in metallicity gradient. The mean log(age/Gyr) gradient is --0.02 $\pm$ 0.06 dex$^{-1}$, although several galaxies have significant positive or negative age gradients. The mean gradient in $\alpha$-element enhancement is --0.10 $\pm$ 0.04 dex$^{-1}$.

We find that stellar population gradients are primarily related to the central metallicity: early-type galaxies with super-solar centres have steep negative metallicity gradients and positive age gradients; those with solar metallicity centres have negligible [Z/H] gradients and negative age gradients. There is a strong observed anti-correlation between the gradients in age and metallicity. While a part of this trend can be attributed to the correlation of measurement errors, we demonstrate that there is an underlying intrinsic relation. For the Shapley galaxies, $B$--$R$ colour gradients predicted from spectroscopic age and metallicity generally agree well with those measured directly from photometry.
\end{abstract}


\section{Introduction}
\label{sec:intro}

For almost four decades, observations have identified systematic radial trends in the stellar properties of galaxies \citep{san72-21}. These internal gradients are a valuable probe into the star formation histories and chemical evolution of galaxies. Traditionally, the focus has been on radial trends in (mostly optical) broadband colours \citep[e.g.][]{pel90-1091,tam00-2134,mic05-451}.

Early-type galaxies typically become steadily bluer towards their outer regions. These colour gradients have often been interpreted as variations in metallicity: \citet{tam03-596} estimated that $\nabla$$B$--$R$ $\sim$ --0.1 corresponds to $\nabla$[Z/H] $\sim$ --0.3\footnote{All gradients are quoted per decade radius, with $\nabla$$X$ the shorthand for the change in X per decade radius}. The relative shallowness of these derived metallicity gradients rule out the simplest models of classic monolithic collapse \citep{lar74-585}. Within the framework of the hierarchical $\Lambda$CDM cosmology, competing processes influence the gradients. For instance, a galaxy with a history of merging would show diluted or erased metallicity gradients \citep{whi80-1}, but a central star burst could strengthen the negative metallicity gradient and create a positive age gradient \citep{bar91-65}.

The well-known degeneracy between age and metallicity in broadband optical colours has hindered attempts to distinguish their respective contribution to the gradients. Using optical and infrared colours to estimate metallicity and age gradients, \citet{tam04-617} found that including the age effect weakens the mean $\nabla$[Z/H] (to --0.16 $\pm$ 0.09), while predicting small age gradients. \citet{wu05-244} used optical/IR colour gradients of 36 nearby early-types to derive a mean metallicity gradient of --0.25 $\pm$ 0.03 and a shallow mean $\nabla$log(age) = 0.02 $\pm$ 0.04, but with a substantial dispersion in age gradients of 0.25.

Spectroscopic data can be used to discriminate more clearly between the age and metallicity effects \citep{wor94-107}. To date, most spectroscopic studies of the radial trends of early-type galaxies have used a single slit along the major axis (e.g. \citealt*{gor90-217}; \citealt{meh03-423,san07-759,red07-1772}), occasionally supplemented by a perpendicular slit along the minor axis (e.g. the S0 study in \citealt*{nor06-815}). Long-slit spectrographs have good optical throughput, but a significant fraction of the galaxy light goes undetected, and the major axis is assumed to be representative of the whole.

An integral field unit (IFU) couples the diagnostic power of spectroscopic observations, to the spatial sampling enjoyed in imaging. The SAURON project, introduced in \citet{bac01-23} and \citet{dez02-513}, observed 72 nearby galaxies ($cz$ $<$ 3000 km s$^{-1}$). The excellent spatial coverage of the instrument permitted two-dimensional dynamical/kinematic mapping of the targets. However, the trade-off for the spatial resolution is a limited spectral range covering only three useful absorption line indices (H$\beta$, Fe5015, Mgb5177), hindering the detailed analysis of stellar populations. \citet{kun06-497} showed that while the spatial maps of H$\beta$ suggest that early-type galaxies typically exhibit no radial trend in age, Fe5015 indicates that metallicity gradients are generally negative (mean $\nabla$Fe5015 = --0.014).

\citet[R08]{raw08-1891} employed the Very Large Telescope (VLT) VIsible MultiObject Spectrograph (VIMOS) IFU to study galaxies in the local cluster Abell 3389 ($cz$ $=$ 8100 km s$^{-1}$). The VIMOS field-of-view is well matched to the size of early-type galaxies at $z$ = 0.02 -- 0.05, and the instrument permits a large optical wavelength range covering many useful absorption line indices. Rather than creating two-dimensional maps of galaxies, radial trends of the spectroscopic features were measured out to the effective radius. From these, gradients were derived in the stellar population parameters (age, metallicity and $\alpha$-element abundance). Generally, galaxies displayed negative metallicity gradients and flat age gradients.

The present paper expands the sample of R08 with new observations of 19 early-type galaxies in the Shapley Supercluster. The Abell 3389 targets have been re-analysed using the improved techniques described below.

This paper adopts the cosmological parameters ($\Omega_{\rm M}$,$\Omega_{\Lambda}$,$h$) = (0.3,0.7,0.7). At the redshift of the Shapley Supercluster ($z$ = 0.048), 1 arcsec corresponds to 0.95 kpc.

\section{Observations and data reduction}
\label{sec:data}

\subsection{Sample and observations}
\label{sec:sample}

The sample consists of 19 early-type galaxies located in the three main constituent clusters (Abell 3556, Abell 3558, Abell 3562) of the Shapley Supercluster core ($<$$cz$$>$ $\approx$ 14660 km s$^{-1}$). Shapley is the richest supercluster in the nearby universe \citep{ray89-251}, and there is a wealth of ancillary data for each galaxy. Photometric coverage includes ESO 2.2m $B$- and $R$-band (Shapley Optical Survey; SOS; \citealt{mer06-109}) and GALEX ultraviolet \citep{raw08-2097}. Inclusion in the high-S/N multi-object spectroscopy of \citet*{smi07-1035} was a prerequisite for selection, and thirteen of the targets also have spectra from the NOAO Fundamental Plane Survey \citep[NFPS;][]{smi04-1558}.

The sample galaxies are all confirmed supercluster members, and were selected to probe a range of masses and age/metallicity combinations (as derived from previous central spectroscopy), and span three magnitudes in luminosity ($m_R$ $\sim$ 14 -- 17 mag; $M_R$ $\sim$ --22.5 -- --19.5 mag, using a distance modulus of 36.5). The galaxies all lie on the $B$--$R$ colour--magnitude relation, and do not exhibit unusually blue $NUV$--$IR$ colours \citep{raw08-2097}. For consistent galaxy-to-galaxy comparisons, the PSF-corrected effective radius ($r_{\rm e}$) is a useful scale-length. We calculate $r_{\rm e}$ from 2MASS imaging, which is available for all our Shapley and Abell 3389 galaxies (see R08 for further details). The Shapley galaxies have effective radii of 1.7 -- 9.7 arcsec (1.6 -- 9.2 kpc). Table \ref{tab:data} lists various parameters for the sample galaxies.

\begin{table*}
\centering
\caption{Observed parameters of the Shapley Supercluster target galaxies. ID from \citet*[][MGP]{met94-431}. Positions, central optical photometry (5 arcsec radius apertures) and morphology (by visual inspection) from SOS $B$- and $R$-band imaging. $J$-band (total) magnitude and PSF-corrected circularised effective radius ($r_{\rm e}$) from 2MASS images. $T_{\rm tot}$ is the co-added VIMOS exposure time. Central S/N \AA$^{-1}$ (averaged over the range 4700--5000 \AA{}) and central velocity dispersion ($\sigma$ km s$^{-1}$; as measured in Section \ref{sec:initial}), both from an $r_{\rm e}$/3 aperture. The final column indicates the number of radial bins used for the derivation of gradients (within $r_{\rm e}$), if the analysis was possible.} 
\label{tab:data} 
\begin{tabular}{@{}cccrrrrcrrrc} 
\hline
ID & RA & Dec & \multicolumn{1}{c}{$m_R$} & \multicolumn{1}{c}{$B$--$R$} & \multicolumn{1}{c}{$m_J$} & \multicolumn{1}{c}{$r_{\rm e}$} & \multicolumn{1}{c}{Morph.} & \multicolumn{1}{c}{$T_{\rm tot}$} & \multicolumn{1}{c}{S/N} & \multicolumn{1}{c}{$\sigma$} & \multicolumn{1}{c}{Bins} \\
(MGP) & (J2000) & (J2000) & \multicolumn{1}{c}{mag} & \multicolumn{1}{c}{mag} & \multicolumn{1}{c}{mag} & \multicolumn{1}{c}{arcsec} & & \multicolumn{1}{c}{(s)} & \multicolumn{1}{c}{\AA$^{-1}$} & \multicolumn{1}{c}{(km s$^{-1}$)} & \# \\
\hline
MGP0129 & 13:24:06.8 & --31:44:49 & 16.42 & 1.37 & 14.62 & 1.96 & E/S0 & 6132 & 14.1 & 103.5 $\pm$ 6.1 &  \\
MGP1189 & 13:26:55.9 & --31:24:47 & 15.46 & 1.47 & 13.50 & 1.66 & E/S0 & 4088 & 44.4 & 289.4 $\pm$ 4.7 & 4 \\
MGP1195 & 13:26:56.0 & --31:25:27 & 15.76 & 1.42 & 13.65 & 3.25 & S0 & 2044 & 22.3 & 115.5 $\pm$ 3.1 & 3 \\
MGP1211 & 13:26:58.3 & --31:30:30 & 16.83 & 1.39 & 15.19 & 1.14 & E & 4088 &  11.7 & 88.9 $\pm$ 9.6 &  \\
MGP1230 & 13:27:00.3 & --31:21:05 & 16.16 & 1.30 & 14.41 & 2.07 & E/S0 & 4088 &  19.7 & 57.5 $\pm$ 3.8 & 3 \\
MGP1440 & 13:27:27.0 & --31:41:07 & 15.92 & 1.43 & 13.97 & 2.34 & E & 6132 & 25.9 & 187.6 $\pm$ 5.5 & 3 \\
MGP1490 & 13:27:33.5 & --31:25:26 & 16.40 & 1.41 & 14.24 & 3.65 & E & 8176 & 14.5 &  76.8 $\pm$ 4.8 & 3 \\
MGP1600 & 13:27:46.6 & --31:27:15 & 15.02 & 1.49 & 12.96 & 2.23 & S0 & 2044 & 43.5 & 297.4 $\pm$ 4.8 & 5 \\
MGP1626 & 13:27:48.5 & --31:28:46 & 15.35 & 1.55 & 13.33 & 1.66 & S0 & 2044 & 36.8 & 259.5 $\pm$ 6.8 & 4 \\
MGP1835 & 13:28:10.5 & --31:23:10 & 15.19 & 1.53 & 12.93 & 3.68 & E & 2044 & 44.3 & 202.1 $\pm$ 4.3 & 5 \\
MGP1988 & 13:28:27.6 & --31:48:18 & 15.76 & 1.54 & 13.69 & 2.48 & E/S0 & 2044 & 26.9 & 175.8 $\pm$ 5.7 & 4 \\
MGP2083 & 13:28:38.7 & --31:20:49 & 14.76 & 1.65 & 11.86 & 9.62 & E & 4088 & 47.8 & 243.9 $\pm$ 3.2 & 7 \\
MGP2146 & 13:28:47.2 & --31:41:35 & 15.69 & 1.42 & 13.36 & 4.61 & S0 & 4088 & 21.6 & 121.3 $\pm$ 4.0 &  \\
MGP2399 & 13:29:14.5 & --31:39:11 & 16.52 & 1.39 & 14.70 & 2.33 & E & 6132 & 11.8 &  80.5 $\pm$ 7.4 &  \\
MGP2437 & 13:29:20.1 & --31:16:39 & 15.74 & 1.54 & 13.84 & 1.87 & E & 2044 & 29.2 & 186.7 $\pm$ 6.1 & 3 \\
MGP2440 & 13:29:20.7 & --31:32:25 & 14.99 & 1.48 & 12.38 & 6.34 & E & 4088 & 50.1 & 227.0 $\pm$ 3.5 & 6 \\
MGP3971 & 13:33:09.5 & --31:36:12 & 15.77 & 1.51 & 13.84 & 1.93 & E/S0 & 2044 & 21.0 & 182.6 $\pm$ 9.2 & 3 \\
MGP3976 & 13:33:11.7 & --31:40:09 & 15.26 & 1.52 & 13.15 & 2.52 & E/S0 & 2044 & 27.5 & 252.6 $\pm$ 5.3 & 4 \\
MGP4358 & 13:34:08.1 & --31:47:34 & 16.86 & 1.46 & 15.11 &  2.20 & E/S0 & 6132 &  9.0 & 68.0 $\pm$ 9.1 &  \\
\hline 
\end{tabular}
\end{table*} 

If present, nebular emission fills in the Balmer absorption lines, causing H$\beta$ to overestimate the galaxy age. \citet{smi07-1035} confirmed that H$\alpha$ emission is a reliable indicator of contaminating H$\beta$ nebular emission. Central spectroscopy shows that eleven of our galaxies have no trace of H$\alpha$ emission. Of the other eight, only three have $EW$(H$\alpha$) $>$ 0.3 \AA{} (MGP0129 MGP2083, MGP3976), and none have $EW$(H$\alpha$) $>$ 0.5 \AA{}. \citet{smi09-1265} calculate that $EW$(H$\alpha$) $=$ 0.5 corresponds to a maximum H$\beta$ contamination of --0.1 \AA, which translates into an age offset of $<$ 15 per cent. At this low level, estimation of the H$\beta$ emission would be unreliable, so we do not attempt corrections.

The targets were observed in April--July 2008 with VIMOS \citep{lef03-1670} on the ESO VLT Melipal at Paranal. VIMOS was used in Integral Field Unit (IFU) mode with the high resolution, blue grism (dispersion $=$ 0.51 \AA{} pixel$^{-1}$), resulting in a wavelength range of 4150 -- 6200 \AA, spectral resolution of 2.1 \AA{} (FWHM) and a field of view of 27 $\times$ 27 arcsec (0.67 $\times$ 0.67 arcsec per element). Observations were made in dark time, with an average seeing of 0.85 arcsec (FWHM). The hour-long observing blocks contained two 1022 s, spatially-offset exposures, and no separate sky frames. We aimed for a minimum S/N $\sim$ 18 \AA$^{-1}$ at 5000 \AA{} for a 1 arcsec wide annulus at the effective radius ($r_{\rm e}$). To achieve this target for the five faintest objects (R $<$ 16.0), the observing block was executed twice. A potential issue with instrument fringing led to the repetition of several planned observations. Data quality was not in fact impacted, so all of the blocks have been included, increasing the co-added exposure time for some galaxies (detailed in Table \ref{tab:data}).

\subsection{Data reduction}
\label{sec:reduction}

The initial data reduction was broadly similar to the techniques detailed in R08. However, there were some significant differences, the salient points of which are described below.

Cosmic rays were removed from the raw frames using {\sc LACosmic} \citep{dok01-1420}. Reduction to a sky-subtracted and wavelength calibrated datacube was achieved using the ESO common pipeline ({\sc EsoRex}) and the {\sc iraf} routine {\sc imcombine}. VIMOS is comprised of four detectors, so every exposure is split into four quadrants. Each quadrant was flat-fielded and wavelength calibrated separately, using arc lamp exposures from immediately after the target observations. A master sky background spectrum was calculated for each quadrant by taking the median over all the spatial pixels in the outer region (rejecting bad pixels). This sky spectrum was then subtracted from each spatial element, pixel-by-pixel, after normalising to the total flux in the [O{\sc I}] $\lambda$5577 \AA{} sky line. The final datacubes were produced by median combining the exposures in all of the observing blocks, followed by relative flux calibration with reference to a spectroscopic standard star. Approximate relative flux calibration is sufficient for a study of spectroscopic gradients in a sample of galaxies at a common redshift.

\begin{figure*}
\begin{minipage}[t]{42mm}
\includegraphics[viewport=80mm 185mm 170mm 240mm,height=42mm,angle=90,clip]{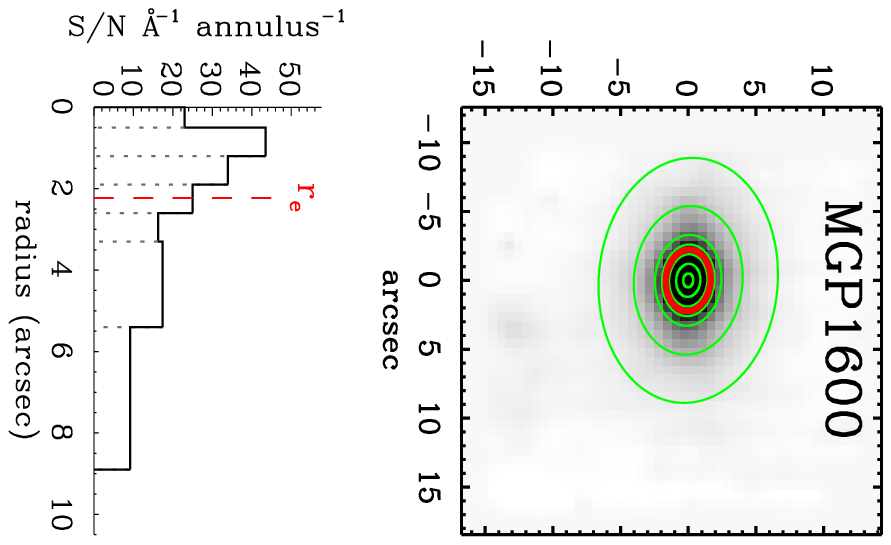}
\end{minipage}
\begin{minipage}[t]{42mm}
\includegraphics[viewport=80mm 185mm 170mm 240mm,height=42mm,angle=90,clip]{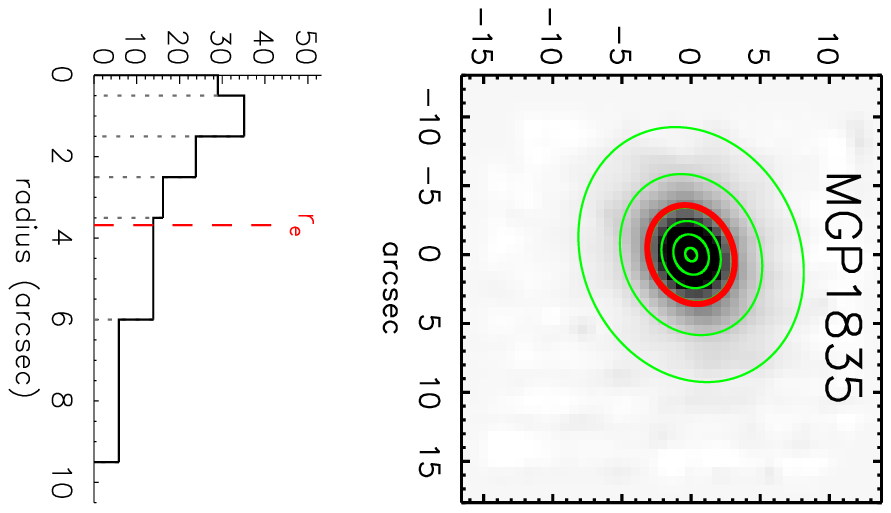}
\end{minipage}
\begin{minipage}[t]{42mm}
\includegraphics[viewport=80mm 185mm 170mm 240mm,height=42mm,angle=90,clip]{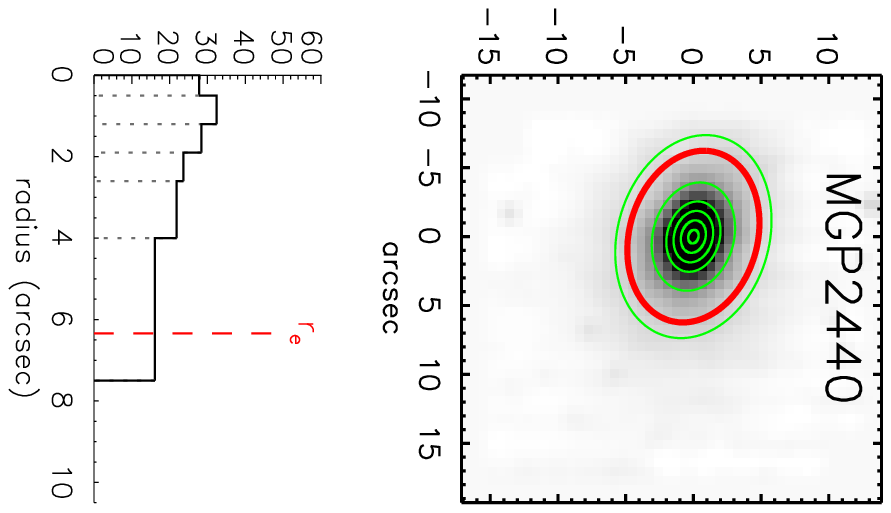}
\end{minipage}
\begin{minipage}[t]{42mm}
\includegraphics[viewport=80mm 185mm 170mm 240mm,height=42mm,angle=90,clip]{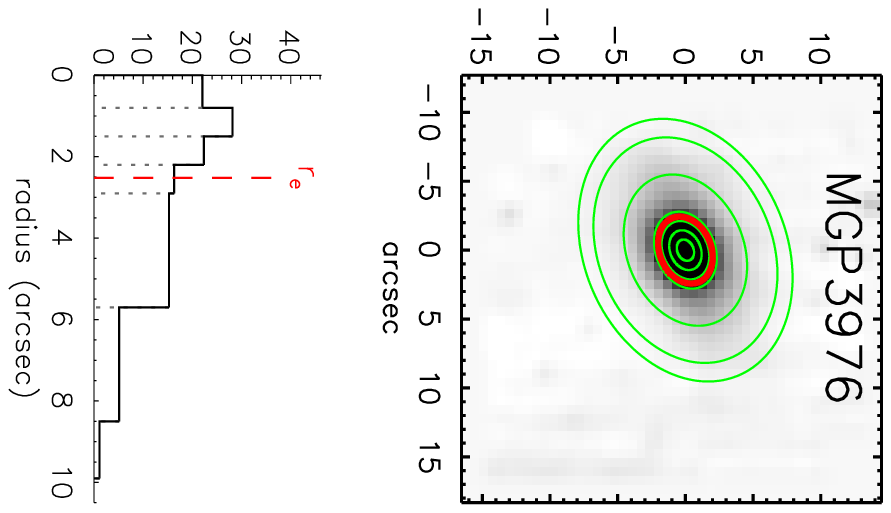}
\end{minipage}
\caption{Adopted radial bins for four example galaxies: (left-to-right) MGP1600, MGP1835, MGP2440, MGP3976. Upper panels show broadband, wavelength-collapsed images from VIMOS, overlaid by the bin boundaries (green) and the effective radius ($r_{\rm e}$; red). Lower panels show the S/N \AA$^{-1}$ (averaged over the region 4700 -- 5000 \AA{}) within each annulus out to 10 arcsec; $r_{\rm e}$ is indicated.}
\label{fig:bins}
\end{figure*}

\subsection{Spatial binning and line index measurements}
\label{sec:initial}

As in \citet{raw08-1891}, the S/N per unit area is too low to attempt full two-dimensional mapping of the spectral features. Instead, a series of annuli are used to compress the data down to one spatial dimension (ellipticity-corrected radius).

The ellipticity and position angle of an ellipse fit to the broadband, wavelength-collapsed image of each galaxy formed the basis of the annuli used as radial bins. The annuli were arranged out to $r_{\rm e}$ for each galaxy individually. Each annulus was required to have a S/N $>$ 18 \AA$^{-1}$ (average for the range 4700--5000 \AA) and a width greater than 0.7 arcsec. Central bins have a minimum semi-major axis of 0.5 arcsec. The galaxies were assigned between three and seven radial bins, depending on their effective radius and surface brightness. As an illustration, the adopted bins for four galaxies are shown in Figure \ref{fig:bins}. The average seeing disc size is similar to the minimum annulus width, so generally each bin is largely independent of neighbouring bins. Spectra were also extracted from within three circular apertures, with radii of 1 arcsec (for direct comparison to previous central spectroscopy; Section \ref{sec:central}), $r_{\rm e}/3$ (for physically consistent central parameters; Section \ref{sec:results}) and $r_{\rm e}$ (used in calculating general aperture corrections; Appendix \ref{sec:app_apcor}).

The velocity dispersion, $\sigma$, was computed from the spectrum of each bin using the fitting routine {\sc ppxf} \citep{cap04-138} and the \citet{vaz99-224} evolutionary stellar population templates. The templates cover a similar wavelength range to VIMOS ($\sim$4900 -- 5500 \AA{} is used for fitting) and have a higher spectral resolution (FWHM = 1.8 \AA). Uncertainties on $\sigma$ were calculated using Monte Carlo simulations.

We used {\sc indexf}\footnote{http://www.ucm.es/info/Astrof/software/indexf/} to measure atomic and molecular Lick indices from de-redshifted, flux-calibrated spectra at the VIMOS instrument resolution ($\sim$50 km s$^{-1}$). For measuring the gradients, model comparisons are performed in a relative sense, so we do not calibrate to the Lick instrumental response. Uncertainties were calculated by {\sc indexf} (as described in \citealt{car98-597}) from error spectra estimated from the variance plane. We correct the measured line indices to account for both the instrument resolution and for broadening due to the observed velocity dispersion of each bin, adopting the same single-stage linear method used in R08. The technique avoids artificially broadening the observed spectra prior to index measurement, preserving any non-uniform noise properties.

\section{IFU central spectroscopy}
\label{sec:central}

One attraction of IFU data is the versatility in possible approaches to the analysis. By extracting a spectrum from a circular aperture, it is trivial to mimic a single-aperture observation. Using the 1 arcsec radius apertures introduced in Section \ref{sec:initial}, the VIMOS data can be compared directly to two previous spectroscopic studies. For reference, these VIMOS central apertures have an average S/N $\sim$ 25 \AA$^{-1}$ over the wavelength range 4700 -- 5000 \AA{}.

The first comparison study is the NOAO Fundamental Plane Survey \citep[NFPS]{smi04-1558}, which analysed 1.5--2 hour observations from the Hydra multi-fibre spectrograph on the 4 m CTIO telescope. The spectral resolution was $\sim$3 \AA{} (FWHM) and the fibre diameter was 2 arcsec. Thirteen of the VIMOS target galaxies were included in NFPS, with an average S/N $\sim$ 40 \AA$^{-1}$ (5000 -- 5500 \AA{}).

The second is a deeper study from \citet{smi07-1035}, which employed 8 hour observations from the AA$\Omega$ fibre-fed multi-object spectrograph on the 3.9 m Anglo-Australian Telescope. The spectral resolution was $\sim$3.5 \AA{} (FWHM) and the fibre diameter was again 2 arcsec. All of the VIMOS targets were observed by the AA$\Omega$ study, with an average S/N $\sim$ 100 \AA$^{-1}$ (for 4700 -- 5000 \AA{}).

\begin{figure}
\includegraphics[viewport=0mm 0mm 138mm 200mm,width=84mm,angle=0,clip]{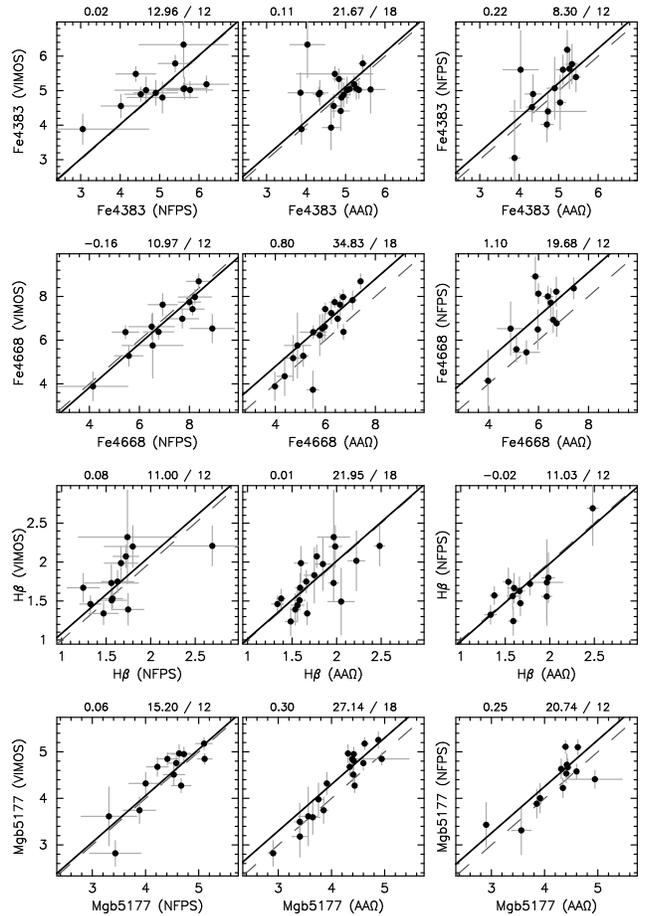}
\caption{Comparison of VIMOS central spectroscopy to NFPS (left column) and AA$\Omega$ (central column). Plots of NFPS versus AA$\Omega$ index measurements for these objects are displayed in the right column. Dashed lines are equality; solid lines shows the offset with the minimum $\chi^2$. The values above each panel show the offset and minimised $\chi^2$ / degrees of freedom.}
\label{fig:line_comp}
\end{figure}

\begin{figure*}
\includegraphics[viewport=0mm 0mm 180mm 55mm,width=165mm,angle=0,clip]{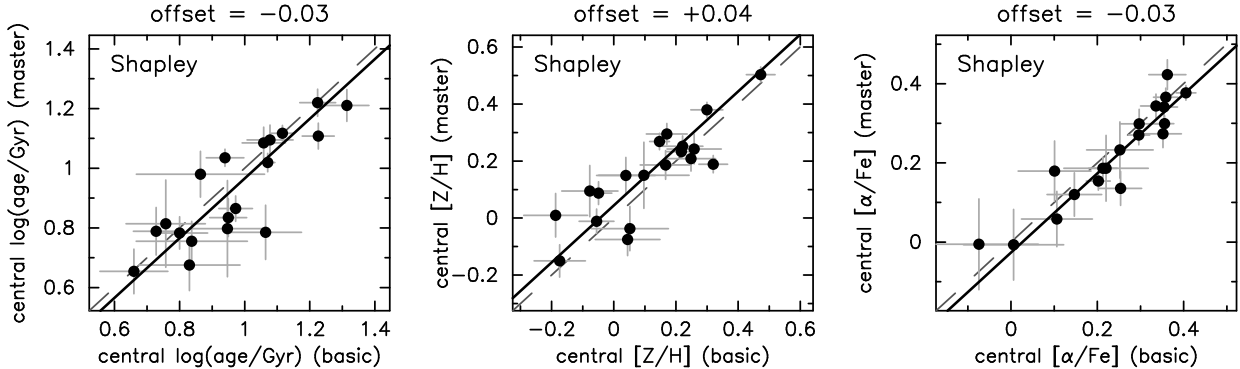}
\caption{Comparison between VIMOS spectroscopic central age (left panel), metallicity (centre panel) and $\alpha$-abundance (right panel) derived using two different index sets: basic (H$\gamma$F + H$\beta$ + Fe4383 + Fe5015 + Mgb5177; top line in Table \ref{tab:ssp_sets}); master (basic + H$\delta$F + Fe4531 + Fe4668 + Fe5406; bold line in Table \ref{tab:ssp_sets}). Solid line shows the weighted mean offset; dashed line is equality. The mean offset is displayed above each panel.}
\label{fig:cent_ssp_sets}
\end{figure*}

\begin{table*}
\centering
\caption{Mean values of the central age, metallicity [Z/H] and $\alpha$-element abundance [$\alpha$/Fe] derived using different index sets in the model inversion. The mean central stellar population parameters derived from AA$\Omega$ observations for the same 19 galaxies, are displayed in the final row for comparison.} 
\label{tab:ssp_sets} 
\begin{tabular}{@{}lrrr} 
\hline
Index set & \multicolumn{1}{c}{$\langle\,$log(age/Gyr)$\,\rangle$} & \multicolumn{1}{c}{$\langle\,$[Z/H]$\,\rangle$} & \multicolumn{1}{c}{$\langle\,$[$\alpha$/Fe]$\,\rangle$} \\
\hline 
H$\gamma$F + H$\beta$ + Fe4383 + Fe5015 + Mgb5177 (\textit{basic}) & 0.97 $\pm$ 0.04 & 0.12 $\pm$ 0.04 & 0.24 $\pm$ 0.03 \\
\textit{basic} + Fe4531 & 0.97 $\pm$ 0.04 & 0.12 $\pm$ 0.04 & 0.24 $\pm$ 0.03 \\
\textit{basic} + Fe5406 & 0.94 $\pm$ 0.04 & 0.14 $\pm$ 0.04 & 0.23 $\pm$ 0.03 \\
\textit{basic} + Fe4668 & 0.95 $\pm$ 0.04 & 0.14 $\pm$ 0.04 & 0.24 $\pm$ 0.03 \\
\textit{basic} + H$\delta$F & 0.97 $\pm$ 0.04 & 0.11 $\pm$ 0.04 & 0.24 $\pm$ 0.03 \\
\textit{basic} + Fe4531 + Fe5406 & 0.95 $\pm$ 0.04 & 0.14 $\pm$ 0.04 & 0.22 $\pm$ 0.03 \\
\textit{basic} + Fe4531 + Fe4668 & 0.95 $\pm$ 0.04 & 0.15 $\pm$ 0.04 & 0.24 $\pm$ 0.03 \\
\textit{basic} + Fe4668 + Fe5406 & 0.94 $\pm$ 0.04 & 0.15 $\pm$ 0.04 & 0.23 $\pm$ 0.03 \\
\textit{basic} + H$\delta$F + Fe4531 + Fe5406 & 0.93 $\pm$ 0.04 & 0.15 $\pm$ 0.04 & 0.22 $\pm$ 0.03 \\
\textit{basic} + H$\delta$F + Fe4531 + Fe4668 & 0.94 $\pm$ 0.04 & 0.14 $\pm$ 0.04 & 0.23 $\pm$ 0.03 \\
\textit{basic} + H$\delta$F + Fe4668 + Fe5406 & 0.93 $\pm$ 0.04 & 0.16 $\pm$ 0.04 & 0.22 $\pm$ 0.03 \\
\textit{basic} + Fe4531 + Fe4668 + Fe5406 & 0.94 $\pm$ 0.04 & 0.16 $\pm$ 0.04 & 0.22 $\pm$ 0.03 \\
\textbf{\textit{basic} + H$\delta$F + Fe4531 + Fe4668 + Fe5406} (\textit{master}) & \textbf{0.93 $\pm$ 0.04} & \textbf{0.16 $\pm$ 0.04} & \textbf{0.22 $\pm$ 0.03} \\
\hline
AA$\Omega$ [VIMOS targets]  \textit{basic} + H$\delta$F & 0.96 $\pm$ 0.04 & 0.10 $\pm$ 0.03 & 0.24 $\pm$ 0.03 \\
\hline
\end{tabular}
\end{table*} 

\subsection{Central line strengths}
\label{sec:line_v_aao}

Figure \ref{fig:line_comp} compares the VIMOS line strength measurements (for Fe4383, Fe4668, H$\beta$ and Mgb5177) to those from the two studies introduced above, using the matched 1 arcsec radius aperture. Agreement between the VIMOS and NFPS values is generally very good, with a reduced $\chi^2$ $\sim$ 1 (after accounting for small mean offsets). Measurements of the Fe4383 and H$\beta$ lines agree well across all three instruments. Fe4668 and Mgb5177 exhibit comparatively large offsets between AA$\Omega$ and both other studies, with the AA$\Omega$ results exhibiting weaker measured lines.

For Fe4668, there is a ($> 5\,\sigma$) correlation between the VIMOS--AA$\Omega$ offset and the index value itself. The Fe4668 index is one of the most susceptible to errors in the continuum level, as it is a particularly broad feature. Exploring factors that could systematically affect the continuum level, we find no correlation between the offset and either airmass or position in the AA$\Omega$ multi-object spectrograph field. Aperture `smearing' (the increase of the effective aperture size in co-added multi-object observations due to fibre placement or telescope pointing errors) may explain a portion of the offset, but cannot account for the entire correlation. While the responsible factor has not been identified, the agreement between VIMOS and NFPS measurements, and their similar offsets from AA$\Omega$, suggests that the problem may lie with the multi-object spectra from AA$\Omega$.

\begin{figure*}
\includegraphics[viewport=0mm 0mm 180mm 55mm,width=165mm,angle=0,clip]{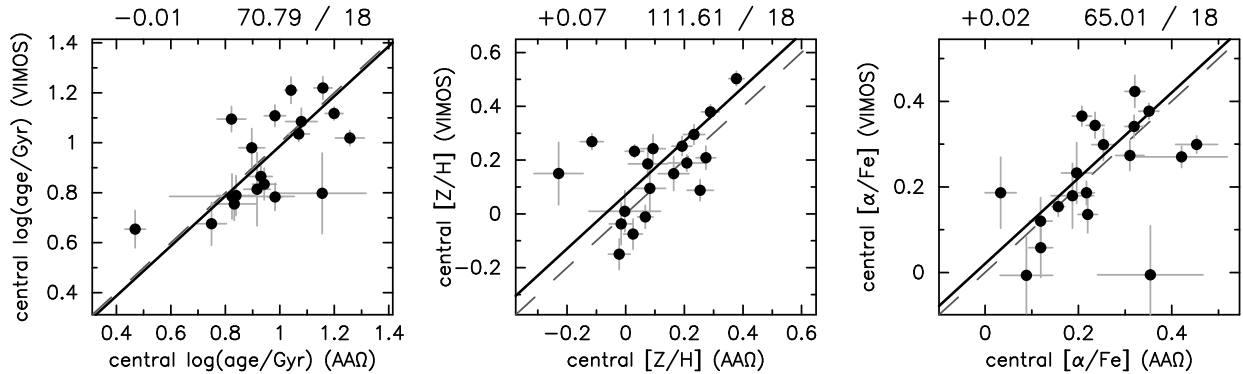}
\caption{Comparison between derived stellar population parameters from VIMOS (using master index set) and AA$\Omega$. Layout as in Figure \ref{fig:cent_ssp_sets}. The distributions and mean values in each parameter are well matched between the two observations, although the errors appear slightly underestimated, particularly in [Z/H].}
\label{fig:cent_ssp}
\end{figure*}

\subsection{Central stellar populations}
\label{sec:ssp_comp}

Stellar populations, characterised by age, total metallicity (Z/H) and $\alpha$-element over-abundance ($\alpha$/Fe), were derived through comparison of the measured line indices to predictions from single burst evolution models. We use a grid inversion technique similar to that of \citet*{pro04-1327} and \citet*{smi09-sub}. This involves a non-linear $\chi^2$ minimisation over a set of indices, using a cubic-spline interpolation in age--[Z/H]--[$\alpha$/Fe], and a linear extrapolation for points lying outside the model grids. We compare to the models of \citet*{tho03-897} and \citet*{tho04-19}, hereafter TMBK.

Nine indices were used in the inversion (the `master' set): three Balmer lines (H$\delta$F, H$\gamma$F, H$\beta$), five iron lines (Fe4383, Fe4531, Fe4668, Fe5015, Fe5406) and the $\alpha$/Fe sensitive Mgb5177. H$\delta$A and H$\gamma$A are considered redundant as they mostly duplicate the response of H$\delta$F and H$\gamma$F. Iron lines towards the red end of the VIMOS wavelength range (Fe5270, Fe5335, Fe5709, Fe5782) are excluded due to contamination by sky features, such as the [O{\sc I}] $\lambda$5577 line.

To test the stability of the inversion technique, we derived central stellar population parameters using several different sets of indices. Starting from a `basic' set of five indices (H$\gamma$F, H$\beta$, Fe4383, Fe5015, Mgb), thirteen sets were created by adding combinations of other lines in the `master' list. The mean characteristics of the 19 galaxies are robust with respect to the different index sets (Table \ref{tab:ssp_sets}). The mean age of the sample varies by less than 5 per cent, and the mean metallicity and $\alpha$-abundance deviate by only 0.04 dex and 0.02 dex respectively. The mean central parameters of the same galaxies derived from AA$\Omega$ spectroscopy (final row, Table \ref{tab:ssp_sets}) are well matched to the VIMOS values. Individually, the derived central stellar populations of galaxies are also independent of the index set used. Figure \ref{fig:cent_ssp_sets} compares values computed from grid inversions using the basic and master sets.

For the remainder of this study, we use the master set of indices (bold row in Table \ref{tab:ssp_sets}) to calculate stellar population parameters. Figure \ref{fig:cent_ssp} compares the central values of age, metallicity and $\alpha$-abundance derived from VIMOS to the equivalent values from AA$\Omega$. Although the mean value for each parameter is well matched between observations, the reduced $\chi^2$ is $\sim$ 4 for age and [$\alpha$/Fe], and $\sim$ 6 for metallicity. It appears that the formal errors, propagated through the index grid inversion by calculating the full error ellipsoids from the $\chi^2$ surface, are underestimated. Errors on the indices themselves are well determined, as noted in Section \ref{sec:line_v_aao} (Figure \ref{fig:line_comp}). Consequently, in the following sections we quote the uncertainty on stellar population parameters as derived from Monte Carlo simulations of the index values.

\section{Gradients}

\subsection{Line strength gradients}
\label{sec:line_grads}

Line strength gradients were measured using an unweighted linear least-squares fit to all the elliptical bins within the effective radius ($r_{\rm e}$; see Section \ref{sec:initial}), such that $\nabla$I = $\Delta$I / $\Delta$log($r/r_{\rm e}$), where I is any line strength index. Most previous studies of radial trends have used the log($r/r_{\rm e}$) convention, resulting in straight forward comparisons. A small fraction of galaxies would be better fit in linear radial space; a weighted fit to these in log($r/r_{\rm e}$) would bias the overall gradient towards the radial trend of the inner region, where the index values are better determined. The unweighted fit ensures a meaningful parametrization of the average gradient out to $r_{\rm e}$ in all galaxies.

Five galaxies (MGP0129, MGP1211, MGP2146, MGP2399, MGP4258) have S/N $<$ 18 \AA$^{-1}$ at $r_{\rm e}$, due to their lower surface brightness. These galaxies are excluded from further analysis as the quality of the spectra is insufficient to calculate reliable gradients.

\begin{figure*}
\includegraphics[viewport=0mm 0mm 175mm 105mm,width=160mm,angle=0,clip]{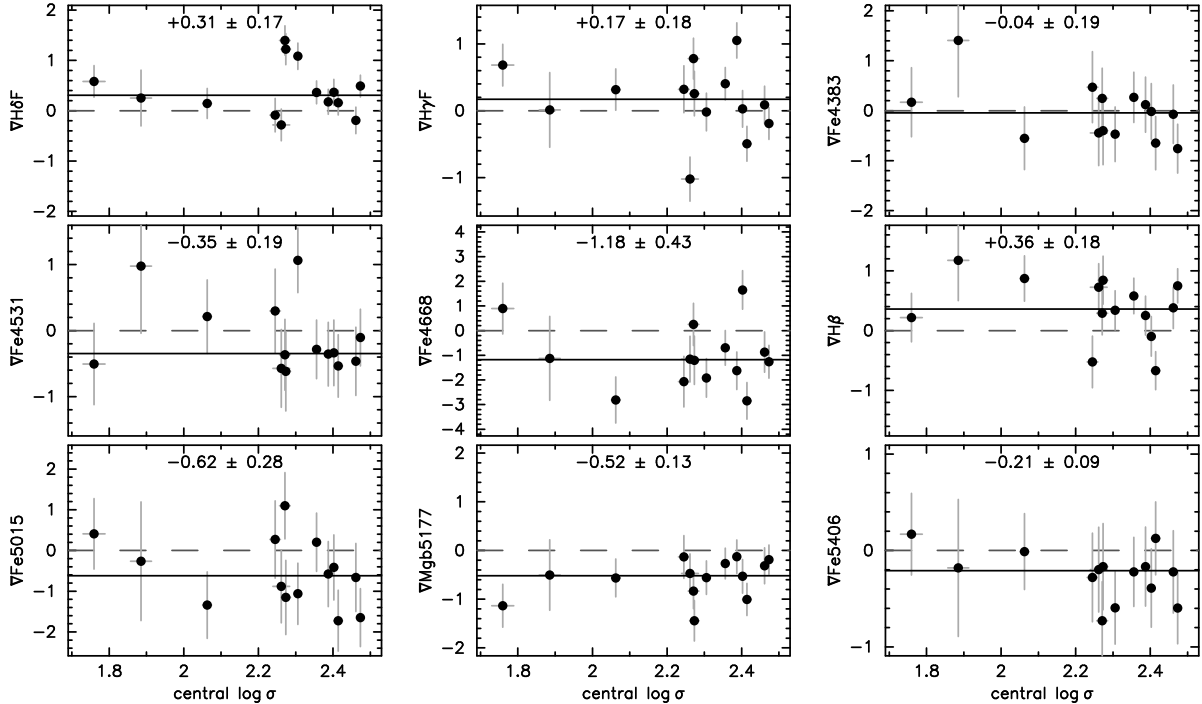}
\caption{Absorption line strength gradients versus central log$\,\sigma$ ($r_{\rm e}$/3 radius aperture). The solid line represents the median gradient (also displayed at the top of each panel); dashed line indicates zero gradient.}
\label{fig:index_grads}
\end{figure*}

Generally, the metal lines show negative gradients, i.e. weaker lines at larger radii, while the Balmer lines have small positive gradients (Figure \ref{fig:index_grads}). The gradients do not exhibit simple trends with the central velocity dispersion, although the exclusion of most of the small galaxies, due to low surface brightness, results in a poor sampling of the low mass end (log$\,\sigma$ $<$ 2.2).

\subsection{Stellar population gradients}
\label{sec:ssp_grads}

For each of the 14 galaxies with measured gradients, the linear fits described above were extrapolated to estimate the index values at $r$ = 0.1$\times$$r_{\rm e}$ and $r$ $=$ $r_{\rm e}$. The corresponding stellar population parameters for these two points were derived using the grid inversion technique described in Section \ref{sec:ssp_comp}, and the `master' index set. Finally, gradients in age, metallicity [Z/H] and $\alpha$-element enhancement [$\alpha$/Fe] were calculated from the difference between these inner and outer points.

This process differs from the approach used in R08, where stellar population parameters were calculated for each annulus, and gradients were fit in a similar manner to the line indices. To test the sensitivity of the results to the choice of technique, Monte Carlo simulations were computed for the measured line strengths (based on the {\sc indexf} errors), and stellar population gradients were derived for each galaxy using both methods. The new method proved to be more stable, relying on 2--3 times fewer model inversions and not requiring a linear fit to noisy stellar population values. The average simulated gradient for each parameter is method-independent. Errors quoted for the stellar population gradients are computed from these Monte Carlo simulations.

The point spread function (PSF) of the observations may lead to an underestimation of the gradients in the smallest galaxies, as central light is smoothed into neighbouring bins. To quantify this effect, we constructed a simple two dimensional model with a known metallicity gradient convolved with an $r^{1/4}$ luminosity profile. Seeing conditions were simulated by smoothing the model with Gaussian filters of widths in the range of the measured PSFs (mean FWHM = 0.85 arcsec; maximum FWHM = 1.2 arcsec). For each smoothed model, the `observed' gradients were derived from radial bins similar to those used for the real data. In the worst realistic case, with PSF FWHM $=$ 0.9$\times$$r_{\rm e}$, the measured gradient was flatter than the actual gradient by 25 per cent. For the majority of the Shapley observations, the effect is estimated to be at the 5 -- 10 per cent level, so no corrections have been applied.

On average, the galaxies in the Shapley sample have a negative gradient in all three stellar population parameters. The mean age gradient is --0.09 $\pm$ 0.07, although there are individual galaxies with significant gradients in either direction. Both metallicity and [$\alpha$/Fe] have marginally significant mean negative gradients: $\langle\,\nabla$[Z/H]$\,\rangle$ = --0.11 $\pm$ 0.05 and $\langle\,\nabla$[$\alpha$/Fe]$\,\rangle$ = --0.06 $\pm$ 0.03. The top row of Figure \ref{fig:all_hists} presents the distributions of the gradients for the 14 Shapley galaxies and Table \ref{tab:grads_appendix} in Appendix \ref{sec:app_grads} lists their values.

\subsection{Trends in gradients}
\label{sec:trends}

\begin{figure}
\includegraphics[viewport=0mm 0mm 165mm 155mm,height=84mm,angle=270,clip]{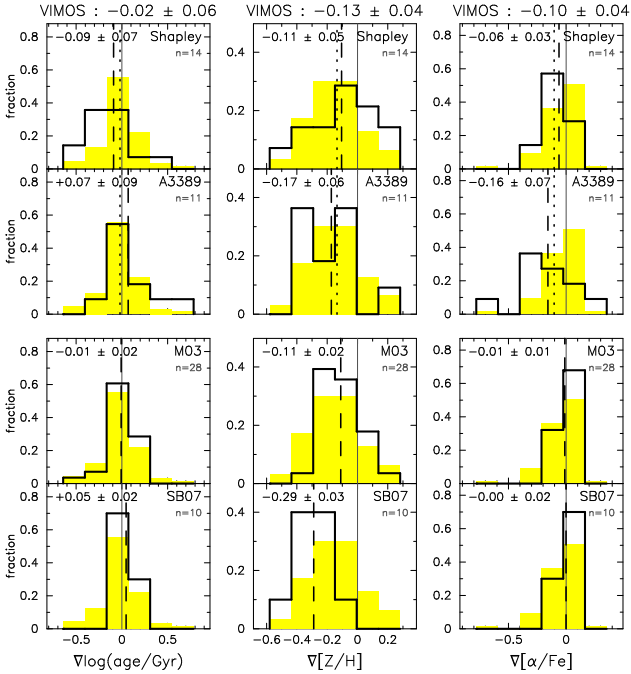}
\caption{Histograms showing the fractional distribution of gradients in (left-to-right) log(age/Gyr), [Z/H] and [$\alpha$/Fe] for galaxies in each sample: (top-to-bottom) Shapley, A3389 (re-analysed R08 data), M03, SB07 (see Section \ref{sec:comp}). Mean gradients for each sample are marked by dashed lines, and are shown in the top--left of each panel. The mean gradients for the VIMOS (Shapley+A3389) sample are marked by a dotted line in the top two rows, and presented at the top of each column. Background shaded histograms are for the overall distribution (Shapley+A3389+M03+SB07).}
\label{fig:all_hists}
\end{figure}

This section presents the correlations between the stellar population gradients and central parameters. The latter are measured from the $r_{\rm e}$/3 apertures, as listed in Table \ref{tab:cent_appendix}. Physical interpretation of the observed dependencies is deferred to Section \ref{sec:results}.

The Shapley results will be combined with gradients derived for the galaxies in Abell 3389 \citep[R08]{raw08-1891}. The raw observations from R08 have been re-analysed using the current methodology, producing stellar population gradients for 11 of the 12 galaxies (D53 has insufficient S/N at $r_{\rm e}$). None of the general conclusions in R08 are altered by the re-worked analysis. The second row of Figure \ref{fig:all_hists} presents the re-calculated gradient distributions for the R08 galaxy sample (hereafter, A3389), and the stellar population characteristics are also presented in Appendix \ref{sec:app_a3389}. The combined set of galaxies from Shapley and A3389 (termed VIMOS) have $\langle\,\nabla$log(age)$\,\rangle$ = --0.02 $\pm$ 0.06, $\langle\,\nabla$[Z/H]$\,\rangle$ = --0.13 $\pm$ 0.04 and $\langle\,\nabla$[$\alpha$/Fe]$\,\rangle$ = --0.10 $\pm$ 0.04.

\begin{table*}
\centering
\caption{Summary of the relations between central parameters (within $r_{\rm e}$/3 apertures) and gradients shown in Figures \ref{fig:cs_gm} -- \ref{fig:ca_gm} and/or mentioned in the text. The slope, error on the slope and significance (in parenthesis) are calculated by minimising the residuals in the Y-direction. The columns display trends for A3389, Shapley, VIMOS (= A3389 + Shapley) and the comparison studies (M03 and SB07). Significant trends in the VIMOS sample ($>$ 3$\,\sigma$) are highlighted in bold type. The final row shows the slope for the bisector of the two simple regressions between age and metallicity gradients (Figure \ref{fig:ga_gm}).} 
\label{tab:trends} 
\begin{tabular}{@{}llrrrrr} 
\hline
\multicolumn{1}{c}{X} & \multicolumn{1}{c}{Y} & \multicolumn{1}{c}{A3389} & \multicolumn{1}{c}{Shapley} & \multicolumn{1}{c}{VIMOS} & \multicolumn{1}{c}{M03} & \multicolumn{1}{c}{SB07} \\
\hline
log$\,\sigma$ & $\nabla$log(age) & +0.01 $\pm$ 0.95 (0.0) & +0.43 $\pm$ 0.34 (1.3) & +0.45 $\pm$ 0.33 (1.4) & --0.02 $\pm$ 0.26 (0.1) & --0.11 $\pm$ 0.18 (0.6) \\
log$\,\sigma$ & $\nabla$[Z/H] & --0.17 $\pm$ 0.62 (0.3) & --0.23 $\pm$ 0.25 (0.9) & --0.25 $\pm$ 0.22 (1.1) & --0.13 $\pm$ 0.24 (0.5) & +0.26 $\pm$ 0.22 (1.2) \\
log$\,\sigma$ & $\nabla$[$\alpha$/Fe] & --0.66 $\pm$ 0.78 (0.9) & +0.42 $\pm$ 0.07 (5.6) & +0.20 $\pm$ 0.22 (0.9) & --0.06 $\pm$ 0.13 (0.5) & +0.13 $\pm$ 0.12 (1.1) \\
log$\,\sigma$ & $\nabla$log$\,\sigma$ & --0.13 $\pm$ 0.22 (0.6) & --0.08 $\pm$ 0.08 (1.0) & --0.07 $\pm$ 0.09 (0.8) & +0.04 $\pm$ 0.08 (0.5) & \multicolumn{1}{c}{--} \\
log(age) & $\nabla$log(age) & --0.15 $\pm$ 0.49 (0.3) & --0.49 $\pm$ 0.28 (1.7) & --0.32 $\pm$ 0.26 (1.3) & --0.09 $\pm$ 0.11 (0.8) & --0.20 $\pm$ 0.08 (2.6) \\
log(age) & $\nabla$[Z/H] & +0.23 $\pm$ 0.31 (0.7) & +0.33 $\pm$ 0.20 (1.6) & +0.27 $\pm$ 0.17 (1.6) & +0.19 $\pm$ 0.10 (1.9) & +0.24 $\pm$ 0.11 (2.2) \\
log(age) & $\nabla$[$\alpha$/Fe] & --0.04 $\pm$ 0.42 (0.1) & +0.22 $\pm$ 0.10 (2.1) & +0.10 $\pm$ 0.17 (0.6) & +0.03 $\pm$ 0.06 (0.5) & +0.00 $\pm$ 0.08 (0.0) \\
\textbf{$[$Z/H]} & \textbf{$\nabla$log(age)} & +1.45 $\pm$ 0.94 (1.5) & +0.97 $\pm$ 0.27 (3.6) & +1.09 $\pm$ 0.28 (\textbf{3.9}) & +0.22 $\pm$ 0.21 (1.0) & +0.20 $\pm$ 0.12 (1.7) \\
\textbf{$[$Z/H]} & \textbf{$\nabla$[Z/H]} & --0.72 $\pm$ 0.64 (1.1) & --0.60 $\pm$ 0.22 (2.8) & --0.63 $\pm$ 0.20 (\textbf{3.1}) & --0.59 $\pm$ 0.17 (3.5) & --0.21 $\pm$ 0.16 (1.3) \\
$[$Z/H] & $\nabla$[$\alpha$/Fe] & --0.07 $\pm$ 0.90 (0.1) & +0.27 $\pm$ 0.13 (2.1) & +0.15 $\pm$ 0.23 (0.6) & --0.08 $\pm$ 0.11 (0.8) & +0.07 $\pm$ 0.10 (0.7) \\
$[\alpha$/Fe] & $\nabla$log(age) & --0.72 $\pm$ 0.78 (0.9) & +0.20 $\pm$ 0.90 (0.2) & --0.51 $\pm$ 0.57 (0.9) & +0.09 $\pm$ 0.42 (0.2) & --0.12 $\pm$ 0.44 (0.3) \\
$[\alpha$/Fe] & $\nabla$[Z/H] & +0.04 $\pm$ 0.53 (0.1) & +0.22 $\pm$ 0.64 (0.3) & +0.18 $\pm$ 0.39 (0.5) & --0.05 $\pm$ 0.39 (0.1) & +0.23 $\pm$ 0.57 (0.4) \\
$[\alpha$/Fe] & $\nabla$[$\alpha$/Fe] & +0.24 $\pm$ 0.69 (0.3) & +0.67 $\pm$ 0.29 (2.3) & +0.51 $\pm$ 0.36 (1.4) & --0.01 $\pm$ 0.21 (0.0) & --0.16 $\pm$ 0.31 (0.5) \\
\hline
\textbf{$\nabla$log(age)} & \textbf{$\nabla$[Z/H]} & --0.75 $\pm$ 0.47 (1.6) & --0.72 $\pm$ 0.08 (9.4) & --0.69 $\pm$ 0.10 (\textbf{7.0}) & --0.95 $\pm$ 0.11 (8.6) & --1.33 $\pm$ 0.37 (3.6) \\
\hline
\end{tabular}
\end{table*} 

We explore the existence of relationships between pairs of central parameters and gradients, using a simple linear regression. The gradient is treated as the dependent variable, as initially our interest lies in determining how well the gradients can be predicted from knowledge of the central properties. Fits were calculated without weighting for errors as intrinsic scatter may be significant. Most pairs of parameters do not show a significant trend in the VIMOS data. Table \ref{tab:trends} lists the slope and significance of each. For example, none of the gradients have a simple dependency on central velocity dispersion. (We explore the $\nabla$[Z/H] versus central log$\,\sigma$ relation further in Section \ref{sec:cs_gm}; see Figure \ref{fig:cs_gm}.) Two significant correlations stand out: the metallicity gradient is related to the central metallicity (central upper panel, Figure \ref{fig:c_g}) with a slope of --0.63 $\pm$ 0.20 (significance of 3.1$\,\sigma$); the age gradient is also correlated with the central metallicity (upper panel, Figure \ref{fig:cm_ga}), with a slope of 1.09 $\pm$ 0.28 (3.9$\,\sigma$). Conversely, neither the age gradient nor the metallicity gradient depend significantly on central age (upper-left panel Figure \ref{fig:c_g} and upper panel Figure \ref{fig:ca_gm} respectively).

The final row of Table \ref{tab:trends} refers to the relation between age gradient and metallicity gradient. In this case, it is unclear which parameter is dependent on the other. Hence, the trend is calculated from the bisecting linear relation of the two simple regressions (age on metallicity; metallicity on age). A strong relation is found in the VIMOS sample, with a slope of --0.69 $\pm$ 0.10 (significance of 7.0$\,\sigma$; upper panel, Figure \ref{fig:ga_gm}).

\section{Discussion}
\label{sec:results}

Many factors potentially contribute to the observed trends in stellar population gradients. Hence, there is a risk of reaching a misleading conclusion due to sparse sampling of a complex parameter space. To minimise this possibility, we test the validity of the trends found in Section \ref{sec:trends} by comparing to data from two published early-type galaxy studies (detailed in Section \ref{sec:comp}). In the first instance, the relationship between central velocity dispersion and metallicity gradient is explored (Section \ref{sec:cs_gm}). We then assess the predictive power of the central stellar population parameters in estimating the gradients (Section \ref{sec:c_g}). Section \ref{sec:g_g} investigates the interplay between the gradients in age and metallicity.

\subsection{Comparison studies}
\label{sec:comp}

The two main comparison studies considered are \citet[][hereafter M03]{meh03-423} and \citet[][SB07]{san07-759}. Both used long-slit spectroscopy to derive stellar population gradients, employing similar techniques and models to our analysis.

\citet{meh03-423} derived stellar population gradients for 32 early-type galaxies in the Coma cluster. The authors rebinned the spectra to achieve a S/N $>$ 30 \AA$^{-1}$ per bin out to $r_{\rm e}$, and computed the stellar population parameters using an index--pair iteration technique, with H$\beta$, Mgb5177 and $\langle$Fe$\rangle$ = (Fe5270+Fe5335)/2, in conjunction with the TMBK models. Their original analysis excluded the two cD galaxies (NGC4889, NGC4874) and an S0 (NGC4865), as they lie beyond the model grids, and NGC4876 due to systematically larger errors. We follow suit, leaving 28 galaxies in the sample. M03 defined central values as the average along the major axis inside 0.1$r_{\rm e}$/$\sqrt{1-\epsilon}$, where $\epsilon$ is the ellipticity. The distributions for the gradients are shown in the third row of Figure \ref{fig:all_hists}. 

\citet{san07-759} studied a sample of 11 local early-type galaxies in various environments (field, group, Virgo cluster), rebinning the spectra to achieve a S/N $>$ 50 \AA$^{-1}$ out to 2$r_{\rm e}$. The authors derived stellar population gradients for each galaxy using a $\chi^2$ minimisation technique, with 19 Lick indices and the TMBK models. NGC2865, a merger remnant, contains an extremely young core, and has been excluded from our analysis. For central values, SB07 used an aperture size of 1.5 arcsec $\times$ $r_{\rm e}$/8. The final row of Figure \ref{fig:all_hists} displays the distribution of gradients for the 10 SB07 galaxies.

In general, the average gradients of the four studies are very similar. The Shapley sample has a more steeply negative mean age gradient, and the combined VIMOS sample has a larger mean $\nabla$[$\alpha$/Fe] than either of the long-slit investigations. The mean negative metallicity gradient from SB07 is nearly twice as steep as in the other samples. The authors derive large negative $\nabla$[Z/H] gradients for galaxies in both high and low density environments, so the difference is not due to their inclusion of non-cluster targets.

\subsection{Metallicity gradient versus central log$\,\sigma$}
\label{sec:cs_gm}

\begin{figure}
\includegraphics[viewport=0mm -10mm 130mm 120mm,height=84mm,angle=270,clip]{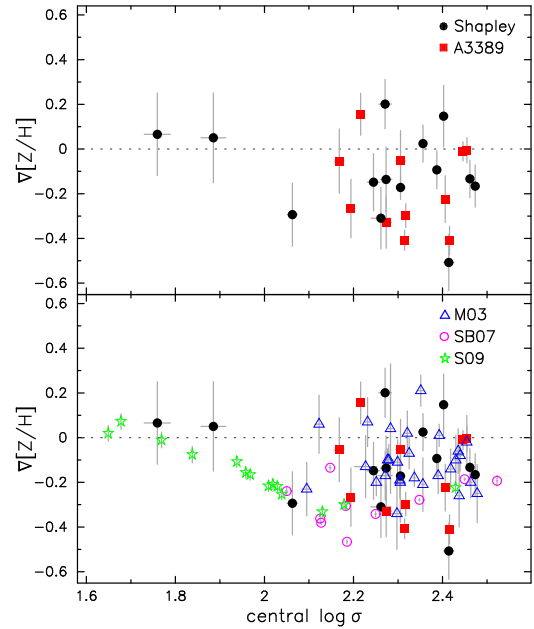}
\caption{Metallicity gradient versus central velocity dispersion (log$\,\sigma$). Upper panel displays the VIMOS data (Shapley = black circles; A3389 = red squares). Lower panel shows VIMOS data together with M03 (blue open triangles), SB07 (magenta open circles) and \citet[S09; green open stars -- measured from their figure 1]{spa09-138}. The two low mass galaxies from Shapley are consistent with the tight relation for low mass early-types in S09.}
\label{fig:cs_gm}
\end{figure}

The combined VIMOS sample does not show a significant correlation between metallicity gradient and the central velocity dispersion (upper panel, Figure \ref{fig:cs_gm}). The lack of a simple linear relationship is confirmed in the comparison studies. As we discuss below, evidence has been previously reported in favour of the [Z/H] gradient depending on mass (via various observable proxies).

A simple model of galaxy formation asserts that enrichment gradients depend on galactic wind efficiency, which scales with the potential and hence galaxy mass \citep{kob04-740}. In a study of 42 early-types, \citet*{car93-553} measured the gradient in the Mg$_2$ absorption line (a simple proxy for metallicity) and discovered that it became shallower with decreasing central velocity dispersion (galaxy mass) for low-mass systems (log$\,\sigma$ $<$ 2.26). High-mass galaxies, however, showed no such correlation. \citet{kob99-573} studied the Mg$_2$ gradients of 80 galaxies (above and below the \citeauthor{car93-553} transition point), but were unable to find a statistically significant trend with any mass estimator (central velocity dispersion, $B$-band luminosity, effective radius or the dynamical mass).

In contrast, \citet*{for05-6} claimed a significant relation between metallicity gradients and galaxy mass ($K$-band luminosity, central velocity dispersion and dynamical mass) for a literature sample of Coma cluster galaxies. Recently, \citet{spa09-138} have resurrected the idea of a mass transition point after observing 14 low mass early-types in nearby clusters (Virgo and Fornax). These authors found that for low mass galaxies (log$\,\sigma$ $<$  2.15), metallicity gradients are remarkably tightly correlated with velocity dispersion, in the sense that (negative) $\nabla$[Z/H] becomes shallower with decreasing log$\,\sigma$ and positive at very low log$\,\sigma$. They conclude that while some higher mass galaxies have steep negative $\nabla$[Z/H], there is also a visible downturn and a markedly broader scatter. Physically, the transition point may mark the mass above which galaxy formation histories are dominated by late-epoch, gas-poor merging, which dilute the metallicity gradient \citep{hop09-135}. The magnitude of the dilution is related to the individual merger properties and the metallicity gradients of the progenitors \citep{dim09-427}, causing the observed inflated scatter at larger masses.

As already noted, Figure \ref{fig:cs_gm} confirms the large scatter and apparent absence of correlation for high mass galaxies. Unfortunately, the A3389, M03 and SB07 studies do not include galaxies below the \citeauthor{spa09-138} mass transition point. Low mass galaxies are also under-represented in the Shapley sample, as several small galaxies have been excluded by low surface brightness at $r_{\rm e}$. However, the remaining galaxies with log$\,\sigma$ $<$ 2.15 are not inconsistent with the mass--metallicity gradient trend reported in \citet{spa09-138}.

\begin{figure*}
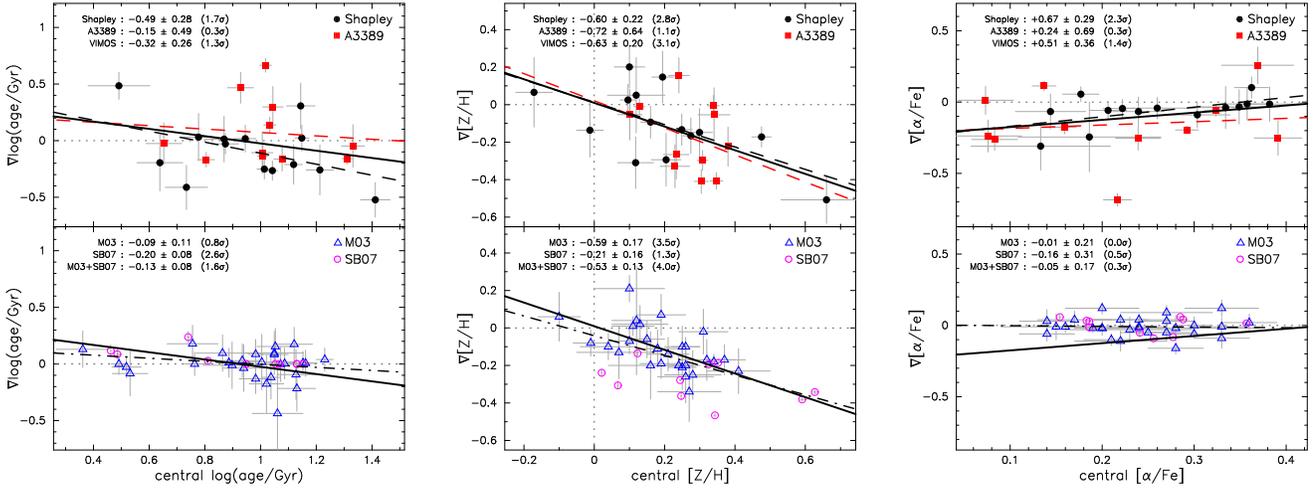

\begin{minipage}[t]{55mm}
\includegraphics[viewport=0mm 0mm 140mm 110mm,height=55mm,angle=270,clip]{figures/ca_ga.eps}
\end{minipage}
\hspace{3mm}
\begin{minipage}[t]{55mm}
\includegraphics[viewport=0mm 0mm 140mm 110mm,height=55mm,angle=270,clip]{figures/cm_gm.eps}
\end{minipage}
\hspace{3mm}
\begin{minipage}[t]{55mm}
\includegraphics[viewport=0mm 0mm 140mm 110mm,height=55mm,angle=270,clip]{figures/cf_gf.eps}
\end{minipage}
\caption{Stellar population gradients versus the corresponding central value for (left-to-right) age, metallicity and $\alpha$-abundance. Upper panels present the VIMOS data (Shapley = black circles; A3389 = red squares), together with best fit lines (dashed; solid black = combined VIMOS) as described in Section \ref{sec:trends}. Lower panels show results from M03 (blue triangles) and SB07 (magenta circles). Dash--dot line indicates the best fit to the combined comparison sample (M03 + SB07); solid line displays the combined VIMOS best fit. The slope and significance of the best fits are displayed to the top-left of the panels, and in Table \ref{tab:trends}.}
\label{fig:c_g}
\end{figure*}

\subsection{Gradient trends with central values}
\label{sec:c_g}

We now explore the relation between central stellar population parameters and gradients. The VIMOS sample shows no significant correlation of the age gradient with central age (upper left panel of Figure \ref{fig:c_g}), or in the analogous relation for $\alpha$-enhancement (upper right panel of Figure \ref{fig:c_g}). The comparison studies confirm the lack of trends in these cases (lower panels), as shown in the final two columns of Table \ref{tab:trends}.

In contrast, the metallicity gradient in VIMOS galaxies is related to the central [Z/H], with a slope of --0.63 $\pm$ 0.20 (3.1$\,\sigma$ significance; upper central panel, Figure \ref{fig:c_g}). This relation is confirmed in the M03 sample, but is not significant in SB07. A similar relation, with a slope $=$ --0.71 $\pm$ 0.34, was reported by \citet{mor08-341} for the bulges of nearby spiral galaxies, which are sometimes considered akin to low-luminosity ellipticals \citep[e.g.][]{tho06-510}. The trend intercepts the origin, and the orientation of the gradients suggest that in general, galaxies tend towards solar metallicities ([Z/H] = 0) at $\sim$ 2--3 $r_{\rm e}$. In rapid dissipative collapse, in-falling gas becomes enriched by evolved stars, and contributes to the metal-rich stars forming in the core. (Recent models by \citealt{pip08-679} predict that as much as 90 per cent of metals in central stars originated in outer regions.) The onset of galactic wind depends on local escape velocity and occurs later in inner regions. As a result, star formation abates earlier in outer regions causing lower [Z/H] at $r_{\rm e}$ than the core \citep{ari87-23,mat94-57}. Both of these processes work to strengthen the metallicity gradient and raise the central Z/H.

The age gradient is correlated with central metallicity in the VIMOS sample, with a slope of 1.09 $\pm$ 0.28 (3.9$\,\sigma$ significance). The relation is such that galaxies with more metal rich centres have stronger age gradients. However, the relation is not significant in either comparison study, or in the A3389 sample alone. The metallicity gradient versus the central age also displays a marginal trend, in which centrally young galaxies favour stronger negative metallicity gradients, and galaxies with old cores exhibit weaker metallicity gradients.

To summarise, the metallicity gradient has a strong dependence on the central metallicity. Although further correlations between parameters are only marginal, some tentative generalisations can be noted. On average, galaxies with metal rich cores have steeper negative metallicity gradients and $\nabla$log(age) $>$ 0. The observed trends are such that galaxies with solar (or sub-solar) metallicity centres tend towards flat metallicity and negative age gradients. These results suggest, or are possibly driven by, a mutual correlation between the age and metallicity gradients.

\begin{figure}
\includegraphics[viewport=0mm -10mm 130mm 120mm,height=84mm,angle=270,clip]{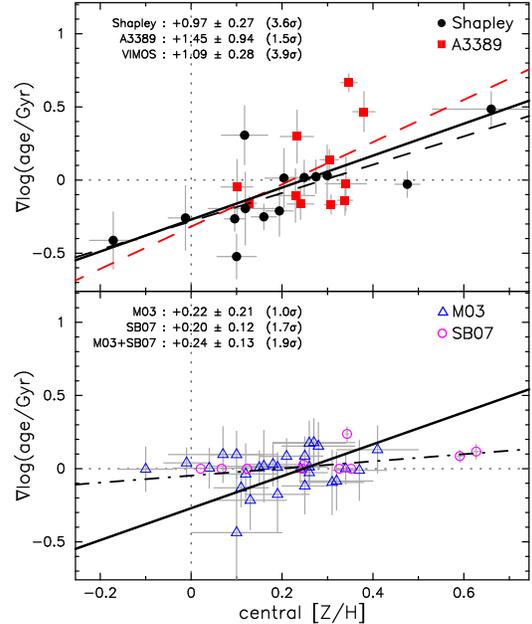}
\caption{Age gradient versus central metallicity. Layout as in each column of Figure \ref{fig:c_g}.}
\label{fig:cm_ga}
\end{figure}

\begin{figure}
\includegraphics[viewport=0mm -10mm 130mm 120mm,height=84mm,angle=270,clip]{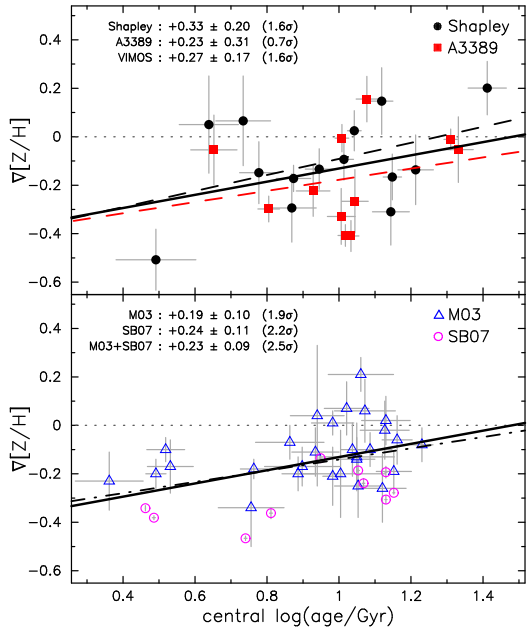}
\caption{Metallicity gradient versus central age. Layout as in each column of Figure \ref{fig:c_g}.}
\label{fig:ca_gm}
\end{figure}

\subsection{Gradient--gradient relation}
\label{sec:g_g}

\begin{figure}
\includegraphics[viewport=0mm -10mm 130mm 120mm,height=84mm,angle=270,clip]{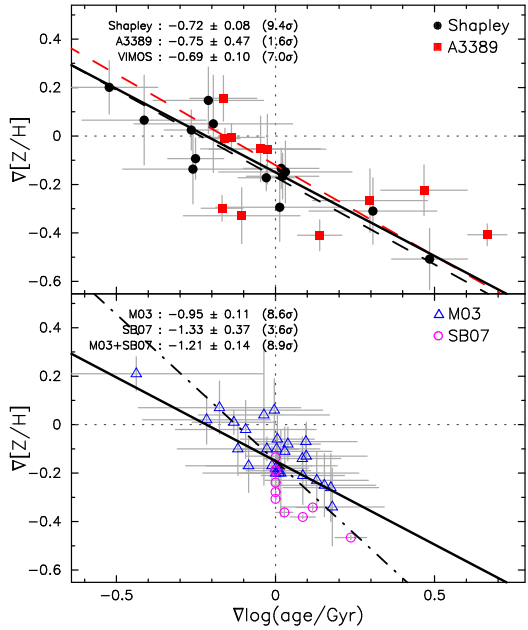}
\caption{Relation between the age and metallicity gradients. Layout as in each column of Figure \ref{fig:c_g}.}
\label{fig:ga_gm}
\end{figure}

The gradients in age and metallicity are strongly anti-correlated in the VIMOS sample, with a slope = --0.69 $\pm$ 0.10 (we discuss the issue of correlated measurement errors in the next paragraph). The relation is confirmed by both comparison studies (Figure \ref{fig:ga_gm}, and the final row of Table \ref{tab:trends}). The trend is reminiscent of the central parameter age--metallicity relation, which also has a strong anti-correlation with a slope of $\sim$0.7 and a significant intrinsic scatter \citep{jor99-607}. However, the central relation is a two-dimensional projection of the three parameter age--mass--metallicity plane (`Z--plane'; \citealt{tra00-165}), with much of the scatter originating from the mass dependence. In contrast, the trend in the gradients has no observed dependence on velocity dispersion.

The strong anti-correlation between $\nabla$log(age) and $\nabla$[Z/H] suggests that generally, a positive age gradient is coupled with a strong negative metallicity gradient, and a negative age gradient is usually associated with a weak or flat metallicity gradient. However, the use of multiple absorption line indices to derive the age and metallicity does not entirely erase the degeneracy, with a correlation dominating the errors (see \citealt{kun01-615} for further discussion). It is prudent to calculate whether the strong trend seen in Figure \ref{fig:ga_gm} could be caused by correlated measurement errors.

\begin{figure}
\includegraphics[viewport=0mm 0mm 180mm 120mm,height=84mm,angle=270,clip]{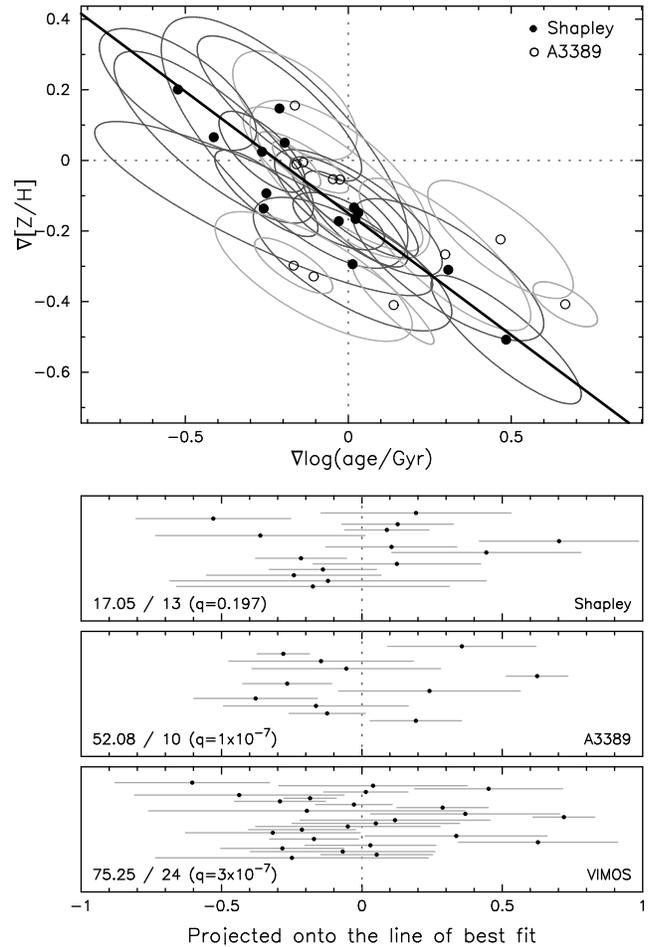}
\caption{Upper panel: The age gradient--metallicity gradient relation for Shapley (solid points) and A3389 (open points), with the error estimates represented by ellipses (Shapley = dark grey; A3389 = light grey). The best fit line is shown. Lower panels: For each sample (top-to-bottom: Shapley, A3389, VIMOS = Shapley + A3389) the points are projected onto the best fit line and shown as a displacement from the median value (y-axis position is arbitrary). The one-dimensional chi-squared, degrees of freedom and q, the probability that the distribution could be entirely attributed to the effect of correlated errors, are displayed.}
\label{fig:ga_gm_ellipse}
\end{figure}

The upper panel of Figure \ref{fig:ga_gm_ellipse} again presents the gradients for the Shapley and A3389 samples. On this occasion, the simple one-dimension (1$\,\sigma$) error bars in each axis have been replaced by the 1$\,\sigma$ error ellipses, based on the distribution of the Monte Carlo realizations. The error ellipses are clearly orientated in the direction of the correlation.

The lower panels of Figure \ref{fig:ga_gm_ellipse} show the distribution of the galaxies projected onto the best-fitting line from the upper panel. We consider a null hypothesis in which the scatter around the zero value is entirely due to the correlated errors. The Shapley sample does not disprove this hypothesis, with an estimated probability of $q$ = 0.2. For the A3389 sample in \citet{raw08-1891}, we concluded that the null hypothesis was incompatible with the data, i.e. that although there is a sizeable correlation of errors, there is also a real intrinsic scatter in the same direction. The newly derived age and metallicity gradients for the A3389 sample confirm this conclusion, as does the combined VIMOS sample. While not contrary to the null hypothesis, the Shapley sample is also not inconsistent with the presence of intrinsic scatter. To explain the age--metallicity gradient relation for the VIMOS sample without invoking intrinsic scatter, the errors in the stellar population gradients would have to be a factor of $\sim$3 larger. Such an underestimation is unlikely, as the gradient errors, derived via Monte Carlo simulations, scale approximately linearly with the well determined index uncertainties.

\subsection{Colour gradients}
\label{sec:colours}

Many studies of broadband colour gradients have ignored the possibility of age gradients (e.g. \citealt{pel90-1091,tam00-2134}), but here we find that $\nabla$log(age) is often inconsistent with zero. By measuring colour gradients in multiple bands, recent studies have attempted to include the age gradient. From SDSS and 2MASS photometry, \citet{wu05-244} derived average $\nabla$[Z/H] = --0.25 $\pm$ 0.03 and $\nabla$log(age) = 0.02 $\pm$ 0.04. However, the age and metallicity predictions are strongly correlated due to the usual degeneracy in broadband colours. For instance, a negative $B$--$R$ colour gradient could be the result of a negative gradient in either metallicity or age. Additionally, our study has found that age and metallicity gradients are intrinsically anti-correlated, affecting the colour gradient in opposite directions.

\begin{figure}
\includegraphics[viewport=0mm 0mm 86mm 82mm,width=84mm,angle=0,clip]{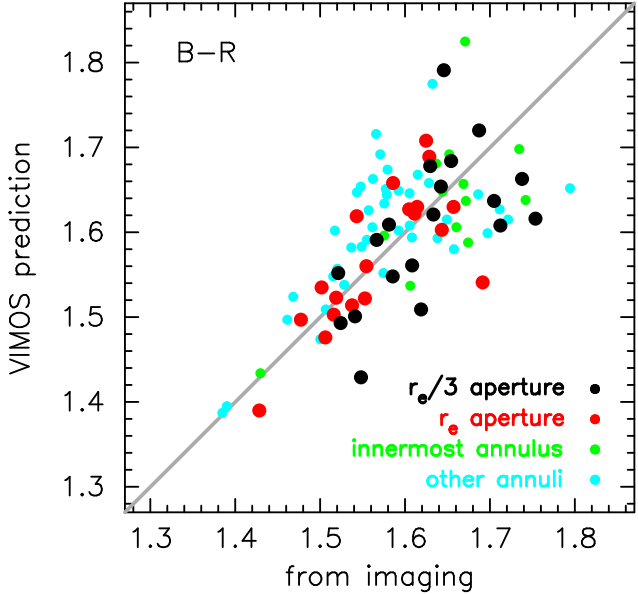}
\caption{Comparison of $B$--$R$ colours from imaging with those derived from the VIMOS stellar population parameters. Each point corresponds to a matched aperture from a target: $r_{\rm e}$/3 radius aperture (black), $r_{\rm e}$ radius aperture (red), innermost annulus from VIMOS radial binning scheme (green), other annuli from VIMOS radial binning scheme (blue). Equality is shown by the solid line. The red ($B$--$R$ $>$ 1.75) predictions from VIMOS are all from MGP2437.}
\label{fig:cols_cent}
\end{figure}

From a known stellar population and given models, the broadband colour can be predicted. In this section, we compare $B$--$R$ colour gradient predictions, calculated from VIMOS spectroscopic age and metallicity, to those observed in $B$- and $R$-band images from the Shapley Optical Survey \citep[][SOS]{mer06-109}. The pixel size is 0.24 arcsec, and the average seeing was $\sim$0.8 arcsec (FWHM) in the $B$-band and $\sim$0.7 arcsec for $R$, corresponding well to the spatial information from VIMOS (0.68 arcsec sampling and an average seeing of 0.85 arcsec). For consistency, the $B$--$R$ gradients were computed using a method as similar as possible to the calculation of VIMOS line strength gradients. $B$- and $R$-band magnitudes were measured for the same elliptical annuli and circular central apertures ($r_{\rm e}$/3 and $r_{\rm e}$) described in Section \ref{sec:initial}. Raw photometry was calibrated onto the Johnson (Vega) system using the SOS image zeropoints, and corrected for galactic extinction with the reddening maps of \citet*{sch98-525}\footnote{A$_{B}$=4.315$\times$$E$($B$--$V$), A$_{R}$=2.673$\times$$E$($B$--$V$)}.

$B$--$R$ colours were predicted from the VIMOS spectroscopic age and metallicity, using the evolutionary stellar population models of \citet[][the basis of the TMBK models]{mar05-799}, which assumes the Kroupa IMF and solar $\alpha$-abundance. These predictions (for individual apertures and annuli) agree well with the photometry, with an overall rms dispersion of 0.068 mag (Figure \ref{fig:cols_cent}). The innermost bins have been distinguished from other annuli, as they are most likely to be affected by the PSF (aperture diameter $\approx$ PSF FWHM).\footnote{The innermost `annuli' are the central ellipses adopted during the spatial binning described in Section \ref{sec:initial}. In several cases they are very similar to the circular $r_{\rm e}$/3 central apertures.}

\begin{figure}
\includegraphics[viewport=0mm 0mm 86mm 82mm,width=84mm,angle=0,clip]{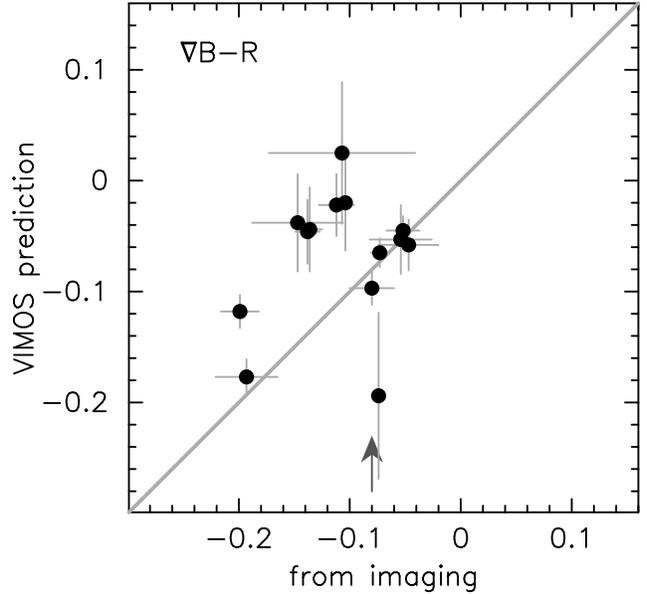}
\caption{Comparison between the $B$--$R$ colour gradients from imaging and VIMOS stellar population parameters. Equality is indicated by the solid line. The arrow displays the mean $\nabla$$B$--$R$ = --0.08 $\pm$ 0.01 mag dex$^{-1}$ from \citet{wu05-244}.}
\label{fig:cols_grad}
\end{figure}

The best VIMOS predictions are for the $r_{\rm e}$ apertures (red points in Figure \ref{fig:cols_cent}), which have high S/N and are not affected by seeing. The comparison has an rms dispersion = 0.056 mag, which is decreased to 0.039 mag if the single outlier (MGP2083: prediction = observed -- 0.16 mag) is excluded. This scatter is consistent with that found by \citet{smi09-sub}, who studied $\sim$230 galaxies in Shapley using AA$\Omega$ spectroscopy and SOS aperture colours. To account for the effect of $\alpha$--abundance on the colour, that study empirically derived the relation $\delta$$B$--$R$ = --0.14 $\times$ $\delta$[$\alpha$/Fe]. Using this, our scatter for the $r_{\rm e}$ apertures is further reduced to 0.034 mag.

The three outlying points with very red predicted colours from VIMOS ($B$--$R$ $>$ 1.75) are all from MGP2437 ($r_{\rm e}$/3 aperture and the two innermost annuli). The red colour prediction results from a large spectroscopic central age estimation (log(age/Gyr) = 1.412; lying beyond the model grids). The extreme age is not observed in the larger 1 arcsec radius aperture, so the cause must be localised in the core. The anomaly appears to be due to extremely weak H$\delta$F and H$\gamma$F measurements, most likely resulting from a combination of poor S/N at the blue end of the spectrum and an unusually deep G4300 line (which lies in the blue `continuum' sideband of H$\gamma$F).

Colour gradients (for both observed and predicted colours) were calculated using an unweighted linear fit to all bins within the effective radius: $\nabla$$B$--$R$ = $\Delta$$B$--$R$ / $\Delta$log($r/r_{\rm e}$). We compared the predicted colour gradients to those observed, as shown in Figure \ref{fig:cols_grad}. The predictions are reasonably consistent with the observations, with a reduced $\chi^2$ $\sim$ 3.4 (rms scatter of 0.08 mag). Accounting for the $\nabla$[$\alpha$/Fe] effect on the colour gradient (using the \citeauthor{smi09-sub} formula above) has only a marginal effect, as the mean correction is $<$ 0.01 mag. The sample mean is well matched to the mean $\nabla$$B$--$R$ = --0.08 $\pm$ 0.01 mag dex$^{-1}$ from \citet{wu05-244}.

\subsection{Towards a physical explanation}

Drawing together the observed trends, we propose the following tentative explanation. Galaxy mass dictates both the efficiency of gas in-fall during rapid dissipative collapse \citep{pip08-679}, and the delayed onset of galactic winds in the core compared to outer regions \citep{mat94-57,kob04-740}. The former process raises the central metallicity, while the latter inhibits metal enrichment at larger radii. Together, they ensure that the strength of the negative metallicity gradient correlates with central metallicity, and that both parameters generally increase with velocity dispersion. For low mass galaxies ($\sigma$ $<$ 140 km s$^{-1}$) these are the prevailing processes, driving the relation for log$\,\sigma$ $<$ 2.15 (e.g. \citealt{car93-553,spa09-138}). Above this transition point, mergers play a dominant role.

Gas-rich merging may prompt a central starburst, resulting in an increased central metallicity and a more negative metallicity gradient \citep{kob04-740}, reinforcing the strong $\nabla$[Z/H]--central [Z/H] relation. The starburst also creates a positive age gradient \citep{mih94-47}, which may be the origin of the $\nabla$log(age)--central [Z/H] relation. However, the logarithmic age gradient from a central starburst is diluted over time, introducing a dependency on the epoch of the last merger, and weakening the $\nabla$log(age) correlations. Late-epoch, gas-poor mergers tend to flatten metallicity gradients \citep{whi80-1,hop09-135}, with the extent of the dilution dependent on the properties of the merger \citep{dim09-427}. This increased scatter in $\nabla$[Z/H] weakens all of the correlations further.

\section{Conclusions}
\label{sec:conclusion}

The internal gradients of early-type galaxies offer a valuable insight into their star formation and chemical evolution. We have used the VIMOS integral field unit to determine the gradients in spectroscopic features for 14 galaxies in the Shapley Supercluster. From these, we derived gradients in age, total metallicity and $\alpha$-element abundance. We combine the Shapley gradients with revised values for our previous VIMOS observations (11 galaxies in Abell 3389; \citealt{raw08-1891}). Our principal results are summarised below.

\begin{itemize}

\item The average metallicity gradient is $\nabla$[Z/H] = --0.13 $\pm$ 0.04.

\item We find that the mean $\nabla$log(age/Gyr) = --0.02 $\pm$ 0.06, but that $\sim$40 per cent of galaxies have age gradients inconsistent with zero ($>$ 2$\,\sigma$).

\item Above a velocity dispersion of log$\,\sigma$ $\sim$ 2.1, galaxies exhibit a large scatter in metallicity gradient, suggesting that mergers with widely varying characteristics dominate their formation histories.

\item Galaxies with metal rich cores have steep negative $\nabla$[Z/H] and have a preference for $\nabla$log(age) $>$ 0. Galaxies with solar (or sub-solar) metallicity centres have $\nabla$[Z/H] $\sim$ 0 and $\nabla$log(age) $<$ 0.

\item There is a strong anti-correlation between the gradients of age and metallicity, with a slope of --0.69 $\pm$ 0.19. We have demonstrated that this relation cannot be attributed to correlated measurement errors.

\item Optical colours and colour gradients derived from the VIMOS spectroscopic age and metallicity are generally consistent with values obtained directly from imaging. Comparison between the spectroscopic and photometric $B$--$R$ gradients has a reduced $\chi^2$ $\sim$ 3.4.

\end{itemize}

A larger set of IFU observations, probing more of the galaxy parameter space (particularly towards lower mass), is required to explore the relationship between the gradients further. Our results have successfully demonstrated the power of the integral field unit in measuring stellar population gradients for local cluster ($z$ $<$ 0.05) early-type galaxies.

\section*{Acknowledgments}
TDR is supported by the STFC Studentship PPA/S/S/2006/04341. RJS is supported by the rolling grant PP/C501568/1 `Extragalactic Astronomy and Cosmology at Durham 2008--2013'. This publication makes use of data products from the Two Micron All Sky Survey, which is a joint project of the University of Massachusetts and the Infrared Processing and Analysis Center/California Institute of Technology, funded by NASA and the National Science Foundation. We also thank Adriana Gargiulo and the SOS team for providing $B$- and $R$- band imaging for the Shapley sample.



\appendix

\section{Aperture correction}
\label{sec:app_apcor}

\begin{table}
\centering
\caption{Median circular aperture corrections I($r_{\rm e}$)$\,-\,$I($r_{\rm e}/3$) for absorption line indices and kinematic/stellar population parameters. Medians, calculated for the 25 galaxies with derived gradients, are in \AA{}, except CN$_1$, CN$_2$, Mg$_1$ and Mg$_2$ which are quoted in mags. Q1 and Q3 give the first and third quartiles of the data.}
\label{tab:app_offset} 
\begin{tabular}{@{}crrr} 
\hline
& \multicolumn{1}{c}{median correction} & \multicolumn{1}{c}{Q1} & \multicolumn{1}{c}{Q3} \\
\hline
H$\delta$A &  0.22 $\pm$  0.31 & --0.38 &  0.74 \\
H$\delta$F &  0.25 $\pm$  0.19 & --0.13 &  0.50 \\
CN$_1$ & --0.02 $\pm$  0.01 & --0.03 &  0.01 \\
CN$_2$ & --0.01 $\pm$  0.01 & --0.04 &  0.00 \\
Ca4227 & --0.18 $\pm$  0.13 & --0.39 &  0.04 \\
G4300 & --0.01 $\pm$  0.16 & --0.48 &  0.28 \\
H$\gamma$A &  0.22 $\pm$  0.17 & --0.36 &  0.57 \\
H$\gamma$F &  0.08 $\pm$  0.11 & --0.11 &  0.44 \\
Fe4383 &  0.13 $\pm$  0.19 & --0.46 &  0.59 \\
Ca4455 & --0.11 $\pm$  0.06 & --0.25 &  0.01 \\
Fe4531 & --0.07 $\pm$  0.11 & --0.32 &  0.17 \\
Fe4668 & --0.81 $\pm$  0.20 & --1.57 & --0.36 \\
H$\beta$ &  0.07 $\pm$  0.08 & --0.06 &  0.27 \\
Fe5015 & --0.12 $\pm$  0.18 & --0.40 &  0.52 \\
Mg$_1$ & --0.02 $\pm$  0.00 & --0.02 & --0.01 \\
Mg$_2$ & --0.02 $\pm$  0.00 & --0.03 & --0.01 \\
Mgb5177 & --0.35 $\pm$  0.08 & --0.59 &  0.01 \\
Fe5270 & --0.27 $\pm$  0.12 & --0.57 & --0.10 \\
Fe5335 & --0.34 $\pm$  0.12 & --0.49 &  0.09 \\
Fe5406 & --0.11 $\pm$  0.07 & --0.25 &  0.14 \\
Fe5709 &  0.01 $\pm$  0.10 & --0.15 &  0.19 \\
Fe5782 & --0.02 $\pm$  0.05 & --0.20 &  0.09 \\
\hline
log$\,\sigma$ & --0.05 $\pm$  0.01 & --0.06 & --0.01 \\
log(age/Gyr) & --0.03 $\pm$  0.05 & --0.18 &  0.05 \\
$[$Z/H$]$ & --0.07 $\pm$  0.04 & --0.19 &  0.04 \\
$[$$\alpha$/Fe$]$ & --0.02 $\pm$  0.03 & --0.10 &  0.02 \\
\hline
\end{tabular}
\end{table} 

Comparison between spectroscopic absorption line strengths from different aperture sizes, requires a correction to account for the gradient. Using the VIMOS IFU data, we have derived aperture corrections in the form I($r_{\rm e}$)$\,-\,$I($r_{\rm e}/3$) for each index, and for log$\,\sigma$, log(age), [Z/H] and [$\alpha$/Fe]. Table \ref{tab:app_offset} presents the median corrections for the 25 galaxies with derived gradients (i.e. excluding D53 from A3389 and MGP0129, MGP1211 MGP2146, MGP2399, MGP4358 from Shapley).

\section{Central value and gradient data}
\label{sec:app_grads}

This paper presents VIMOS observations of 19 galaxies in the Shapley Supercluster core. Table \ref{tab:cent_appendix} lists the central age, total metallicity and $\alpha$-element over-abundace, as derived from spectra observed using an $r_{\rm e}$/3 aperture. Table \ref{tab:grads_appendix} presents the stellar population gradients for 14 of the galaxies. Table \ref{tab:cols_appendix} lists two sets of $B$--$R$ central colours and colour gradients for each galaxy: photometric from imaging; predicted from VIMOS stellar populations.

\begin{table*}
\centering
\caption{Central velocity dispersion and derived stellar populations for the 19 galaxies in the Shapley sample. Central parameters are calculated from an $r_{\rm e}$/3 aperture.} 
\label{tab:cent_appendix} 
\begin{tabular}{@{}lrrrr} 
\hline
 ID & \multicolumn{1}{c}{log$\,\sigma$} & \multicolumn{1}{c}{log(age/Gyr)} & \multicolumn{1}{c}{[Z/H]} & \multicolumn{1}{c}{[$\alpha$/Fe]} \\
 \hline
MGP0129 & 2.015 $\pm$ 0.026 & 0.867 $\pm$ 0.115 & --0.024 $\pm$ 0.061 & 0.090 $\pm$ 0.089 \\
MGP1189 & 2.461 $\pm$ 0.007 & 0.945 $\pm$ 0.039 & 0.249 $\pm$ 0.030 & 0.240 $\pm$ 0.028 \\
MGP1195 & 2.063 $\pm$ 0.012 & 0.869 $\pm$ 0.068 & 0.204 $\pm$ 0.042 & 0.144 $\pm$ 0.038 \\
MGP1211 & 1.949 $\pm$ 0.050 & 1.013 $\pm$ 0.079 & 0.007 $\pm$ 0.101 & --0.016 $\pm$ 0.129 \\
MGP1230 & 1.759 $\pm$ 0.029 & 0.734 $\pm$ 0.076 & --0.171 $\pm$ 0.052 & 0.133 $\pm$ 0.076 \\
MGP1440 & 2.273 $\pm$ 0.013 & 1.213 $\pm$ 0.049 & --0.012 $\pm$ 0.038 & 0.334 $\pm$ 0.037 \\
MGP1490 & 1.885 $\pm$ 0.028 & 0.639 $\pm$ 0.082 & 0.119 $\pm$ 0.082 & 0.186 $\pm$ 0.078 \\
MGP1600 & 2.473 $\pm$ 0.007 & 1.148 $\pm$ 0.026 & 0.274 $\pm$ 0.022 & 0.362 $\pm$ 0.018 \\
MGP1626 & 2.414 $\pm$ 0.012 & 0.491 $\pm$ 0.111 & 0.660 $\pm$ 0.130 & 0.260 $\pm$ 0.042 \\
MGP1835 & 2.306 $\pm$ 0.009 & 0.874 $\pm$ 0.048 & 0.475 $\pm$ 0.028 & 0.177 $\pm$ 0.020 \\
MGP1988 & 2.245 $\pm$ 0.014 & 0.778 $\pm$ 0.054 & 0.299 $\pm$ 0.036 & 0.303 $\pm$ 0.036 \\
MGP2083 & 2.387 $\pm$ 0.006 & 1.014 $\pm$ 0.027 & 0.160 $\pm$ 0.028 & 0.222 $\pm$ 0.021 \\
MGP2146 & 2.084 $\pm$ 0.015 & 0.836 $\pm$ 0.080 & 0.109 $\pm$ 0.044 & 0.225 $\pm$ 0.045 \\
MGP2399 & 1.906 $\pm$ 0.042 & 0.662 $\pm$ 0.098 & 0.073 $\pm$ 0.106 & 0.051 $\pm$ 0.088 \\
MGP2437 & 2.271 $\pm$ 0.014 & 1.412 $\pm$ 0.053 & 0.100 $\pm$ 0.043 & 0.382 $\pm$ 0.035 \\
MGP2440 & 2.356 $\pm$ 0.007 & 1.043 $\pm$ 0.019 & 0.096 $\pm$ 0.022 & 0.206 $\pm$ 0.020 \\
MGP3971 & 2.261 $\pm$ 0.022 & 1.144 $\pm$ 0.051 & 0.117 $\pm$ 0.051 & 0.349 $\pm$ 0.044 \\
MGP3976 & 2.402 $\pm$ 0.009 & 1.118 $\pm$ 0.031 & 0.194 $\pm$ 0.030 & 0.358 $\pm$ 0.025 \\
MGP4358 & 1.832 $\pm$ 0.070 & 0.659 $\pm$ 0.153 & 0.237 $\pm$ 0.148 & --0.188 $\pm$ 0.120 \\
\hline 
\end{tabular}
\end{table*}

\begin{table*}
\centering
\caption{Velocity dispersion and stellar population gradients for 14 galaxies in the Shapley sample.} 
\label{tab:grads_appendix} 
\begin{tabular}{@{}lrrrr} 
\hline
 ID & \multicolumn{1}{c}{$\nabla$log$\,\sigma$} & \multicolumn{1}{c}{$\nabla$log(age/Gyr)} & \multicolumn{1}{c}{$\nabla$[Z/H]} & \multicolumn{1}{c}{$\nabla$[$\alpha$/Fe]} \\
 \hline
MGP1189 & --0.137 $\pm$ 0.040 & 0.018 $\pm$ 0.119 & --0.133 $\pm$ 0.084 & --0.067 $\pm$ 0.102 \\
MGP1195 & --0.094 $\pm$ 0.057 & 0.013 $\pm$ 0.201 & --0.294 $\pm$ 0.141 & --0.066 $\pm$ 0.123 \\
MGP1230 & --0.073 $\pm$ 0.130 & --0.413 $\pm$ 0.195 & 0.066 $\pm$ 0.185 & --0.310 $\pm$ 0.167 \\
MGP1440 & --0.152 $\pm$ 0.058 & --0.259 $\pm$ 0.221 & --0.136 $\pm$ 0.144 & --0.038 $\pm$ 0.148 \\
MGP1490 & --0.038 $\pm$ 0.031 & --0.195 $\pm$ 0.250 & 0.050 $\pm$ 0.201 & --0.245 $\pm$ 0.244 \\
MGP1600 & --0.071 $\pm$ 0.028 & 0.022 $\pm$ 0.114 & --0.166 $\pm$ 0.095 & 0.102 $\pm$ 0.075 \\
MGP1626 & --0.145 $\pm$ 0.046 & 0.484 $\pm$ 0.120 & --0.508 $\pm$ 0.126 & --0.042 $\pm$ 0.074 \\
MGP1835 & --0.088 $\pm$ 0.026 & --0.029 $\pm$ 0.089 & --0.172 $\pm$ 0.054 & 0.055 $\pm$ 0.052 \\
MGP1988 & --0.179 $\pm$ 0.039 & 0.031 $\pm$ 0.209 & --0.148 $\pm$ 0.128 & --0.090 $\pm$ 0.090 \\
MGP2083 & --0.046 $\pm$ 0.029 & --0.251 $\pm$ 0.088 & --0.093 $\pm$ 0.085 & --0.045 $\pm$ 0.064 \\
MGP2437 & --0.232 $\pm$ 0.112 & --0.523 $\pm$ 0.153 & 0.201 $\pm$ 0.110 & --0.014 $\pm$ 0.120 \\
MGP2440 & --0.036 $\pm$ 0.011 & --0.265 $\pm$ 0.085 & 0.025 $\pm$ 0.083 & --0.058 $\pm$ 0.064 \\
MGP3971 & --0.088 $\pm$ 0.010 & 0.306 $\pm$ 0.201 & --0.310 $\pm$ 0.137 & --0.034 $\pm$ 0.121 \\
MGP3976 & --0.198 $\pm$ 0.038 & --0.211 $\pm$ 0.174 & 0.147 $\pm$ 0.138 & --0.014 $\pm$ 0.113 \\
\hline 
\end{tabular}
\end{table*}

\begin{table*}
\centering
\caption{Central $B$--$R$ colour ($r_{\rm e}$/3 aperture) and colour gradient for each galaxy in the Shapley sample. Photometric measurements are from WFI $B$- and $R$-band imaging, in Vega magnitudes and corrected for galactic extinction. Colour gradients are calculated using the same annuli as the spectroscopic gradients. Predicted values are derived from the VIMOS spectroscopic age and metallicity, via the evolutionary stellar population models of \citet{mar05-799}} 
\label{tab:cols_appendix} 
\begin{tabular}{@{}lcrcr} 
\hline
 ID & \multicolumn{2}{c}{Photometric} & \multicolumn{2}{c}{VIMOS predicted} \\
 & \multicolumn{1}{c}{central $B$--$R$} & \multicolumn{1}{c}{$\nabla$$B$--$R$} & \multicolumn{1}{c}{central $B$--$R$} & \multicolumn{1}{c}{$\nabla$$B$--$R$} \\
 \hline
MGP0129 & 1.62 & \multicolumn{1}{c}{--} & 1.51 & \multicolumn{1}{c}{--} \\
MGP1189 & 1.63 & --0.15 $\pm$ 0.04 & 1.62 & --0.04 $\pm$ 0.04 \\
MGP1195 & 1.57 & --0.08 $\pm$ 0.02 & 1.59 & --0.10 $\pm$ 0.01 \\
MGP1211 & 1.61 & \multicolumn{1}{c}{--} & 1.56 & \multicolumn{1}{c}{--} \\
MGP1230 & 1.55 & --0.05 $\pm$ 0.03 & 1.43 & --0.05 $\pm$ 0.03 \\
MGP1440 & 1.64 & --0.19 $\pm$ 0.03 & 1.65 & --0.18 $\pm$ 0.02 \\
MGP1490 & 1.54 & --0.14 $\pm$ 0.01 & 1.50 & --0.04 $\pm$ 0.04 \\
MGP1600 & 1.69 & --0.05 $\pm$ 0.02 & 1.72 & --0.04 $\pm$ 0.01 \\
MGP1626 & 1.70 & --0.11 $\pm$ 0.02 & 1.64 & --0.02 $\pm$ 0.03 \\
MGP1835 & 1.65 & --0.05 $\pm$ 0.03 & 1.68 & --0.06 $\pm$ 0.02 \\
MGP1988 & 1.71 & --0.14 $\pm$ 0.01 & 1.61 & --0.05 $\pm$ 0.03 \\
MGP2083 & 1.75 & --0.20 $\pm$ 0.02 & 1.62 & --0.12 $\pm$ 0.01 \\
MGP2146 & 1.52 & \multicolumn{1}{c}{--} & 1.55 & \multicolumn{1}{c}{--} \\
MGP2399 & 1.52 & \multicolumn{1}{c}{--} & 1.49 & \multicolumn{1}{c}{--} \\
MGP2437 & 1.55 & --0.07 $\pm$ 0.01 & 1.79 & --0.19 $\pm$ 0.07 \\
MGP2440 & 1.58 & --0.07 $\pm$ 0.02 & 1.61 & --0.07 $\pm$ 0.01 \\
MGP3971 & 1.74 & --0.11 $\pm$ 0.07 & 1.66 & +0.03 $\pm$ 0.06 \\
MGP3976 & 1.63 & --0.10 $\pm$ 0.02 & 1.68 & --0.02 $\pm$ 0.04 \\
MGP4358 & 1.59 & \multicolumn{1}{c}{--} & 1.55 & \multicolumn{1}{c}{--} \\
\hline 
\end{tabular}
\end{table*}

\section{Revised results for Rawle et al. (2008b) sample (A3389)}
\label{sec:app_a3389}

The Shapley sample presented in this paper is combined with the galaxies from Abell 3389 introduced in \citet{raw08-1891}. The raw data was re-analysed using the improved techniques described. Particularly important is the new and more stable implementation of the gradient computation, which requires 2--3 times fewer grid inversions and no linear fitting to noisy stellar population data points. Table \ref{tab:cent_r08_appendix} presents the central velocity dispersion and stellar population parameters (using $r_{\rm e}$/3 apertures), while Table \ref{tab:grads_r08_appendix} lists the revised gradients.

\begin{table*}
\centering
\caption{Revised central velocity dispersion and stellar populations for the 12 galaxies in the A3389 sample (observations from \citealt{raw08-1891}). Central parameters are calculated from an $r_{\rm e}$/3 aperture.} 
\label{tab:cent_r08_appendix} 
\begin{tabular}{@{}lrrrr} 
\hline
 ID & \multicolumn{1}{c}{log$\,\sigma$} & \multicolumn{1}{c}{log(age/Gyr)} & \multicolumn{1}{c}{[Z/H]} & \multicolumn{1}{c}{[$\alpha$/Fe]} \\
\hline
D40 & 2.273 $\pm$ 0.012 & 1.006 $\pm$ 0.038 & 0.230 $\pm$ 0.042 & 0.369 $\pm$ 0.039 \\
D41 & 2.315 $\pm$ 0.005 & 1.018 $\pm$ 0.021 & 0.347 $\pm$ 0.018 & 0.137 $\pm$ 0.013 \\
D42 & 2.455 $\pm$ 0.005 & 1.008 $\pm$ 0.019 & 0.339 $\pm$ 0.017 & 0.324 $\pm$ 0.012 \\
D43 & 2.194 $\pm$ 0.009 & 1.044 $\pm$ 0.037 & 0.233 $\pm$ 0.037 & 0.073 $\pm$ 0.030 \\
D44 & 2.406 $\pm$ 0.009 & 0.928 $\pm$ 0.047 & 0.380 $\pm$ 0.025 & 0.239 $\pm$ 0.024 \\
D48 & 2.216 $\pm$ 0.009 & 1.077 $\pm$ 0.030 & 0.241 $\pm$ 0.030 & 0.083 $\pm$ 0.025 \\
D49 & 2.306 $\pm$ 0.011 & 1.331 $\pm$ 0.042 & 0.101 $\pm$ 0.035 & 0.391 $\pm$ 0.026 \\
D52 & 2.446 $\pm$ 0.005 & 1.310 $\pm$ 0.019 & 0.128 $\pm$ 0.016 & 0.292 $\pm$ 0.012 \\
D53 & 2.077 $\pm$ 0.012 & 0.642 $\pm$ 0.072 & 0.354 $\pm$ 0.048 & 0.182 $\pm$ 0.042 \\
D60 & 2.317 $\pm$ 0.003 & 0.805 $\pm$ 0.030 & 0.308 $\pm$ 0.015 & 0.159 $\pm$ 0.013 \\
D61 & 2.168 $\pm$ 0.011 & 0.652 $\pm$ 0.064 & 0.340 $\pm$ 0.045 & 0.076 $\pm$ 0.037 \\
D89 & 2.416 $\pm$ 0.006 & 1.033 $\pm$ 0.022 & 0.304 $\pm$ 0.020 & 0.217 $\pm$ 0.015 \\
\hline 
\end{tabular}
\end{table*}

\begin{table*}
\centering
\caption{Revised velocity dispersion and stellar population gradients for 11 galaxies in the A3389 sample (D53 is excluded).} 
\label{tab:grads_r08_appendix} 
\begin{tabular}{@{}lrrrr} 
\hline
 ID & \multicolumn{1}{c}{$\nabla$log$\,\sigma$} & \multicolumn{1}{c}{$\nabla$log(age/Gyr)} & \multicolumn{1}{c}{$\nabla$[Z/H]} & \multicolumn{1}{c}{$\nabla$[$\alpha$/Fe]} \\
\hline
D40 & 0.075 $\pm$ 0.085 & --0.106 $\pm$ 0.184 & --0.329 $\pm$ 0.115 & 0.258 $\pm$ 0.128 \\
D41 & 0.037 $\pm$ 0.035 & 0.666 $\pm$ 0.062 & --0.407 $\pm$ 0.045 & 0.113 $\pm$ 0.034 \\
D42 & --0.076 $\pm$ 0.008 & --0.138 $\pm$ 0.103 & --0.006 $\pm$ 0.056 & --0.058 $\pm$ 0.041 \\
D43 & --0.072 $\pm$ 0.023 & 0.297 $\pm$ 0.181 & --0.266 $\pm$ 0.131 & 0.012 $\pm$ 0.098 \\
D44 & --0.097 $\pm$ 0.018 & 0.467 $\pm$ 0.136 & --0.224 $\pm$ 0.105 & --0.254 $\pm$ 0.083 \\
D48 & --0.090 $\pm$ 0.049 & --0.164 $\pm$ 0.105 & 0.156 $\pm$ 0.094 & --0.262 $\pm$ 0.076 \\
D49 & --0.036 $\pm$ 0.006 & --0.047 $\pm$ 0.188 & --0.053 $\pm$ 0.135 & --0.252 $\pm$ 0.121 \\
D52 & 0.002 $\pm$ 0.012 & --0.159 $\pm$ 0.048 & --0.011 $\pm$ 0.043 & --0.195 $\pm$ 0.028 \\
D60 & --0.069 $\pm$ 0.011 & --0.168 $\pm$ 0.066 & --0.298 $\pm$ 0.053 & --0.177 $\pm$ 0.047 \\
D61 & --0.046 $\pm$ 0.064 & --0.026 $\pm$ 0.184 & --0.054 $\pm$ 0.144 & --0.238 $\pm$ 0.112 \\
D89 & --0.268 $\pm$ 0.042 & 0.139 $\pm$ 0.070 & --0.410 $\pm$ 0.064 & --0.685 $\pm$ 0.045 \\
\hline 
\end{tabular}
\end{table*}

\label{lastpage}

\end{document}